%% file: Ising_Resetting_0912.tex
\documentclass[a4,10pt]{article}
\usepackage[utf8]{inputenc} 
\usepackage[english]{babel} 
\usepackage{graphicx}
\usepackage{amsmath}
\usepackage{xcolor}
\usepackage{doi}
\usepackage{vmargin}
\setmarginsrb{3cm}{3cm}{3cm}{3cm}{0cm}{0cm}{0cm}{1cm} 

\usepackage{lineno}
\usepackage{enumitem}
\usepackage{amssymb}
\usepackage{amsthm}

\newcommand{\rmc}{{\rm c}}
\newcommand{\rme}{{\rm e}}
\newcommand{\rmd}{{\rm d}}

\newcommand{\ud}[1]{\hspace{-0em}\mathrm{d}{#1}\;}

\newcommand{\Ud}[1]{\hspace{-0ex}\mathrm{d}{#1}\;}

\newtheorem*{theorem}{Claim}

\usepackage[symbol]{footmisc}

\begin{document}
\begin{center}{\Large \textbf{
\hspace{-1cm} Non-Analytic Non-Equilibrium Field Theory:\\
Stochastic Reheating of the Ising Model
}}\end{center}

\begin{center}
Camille Aron\textsuperscript{1}\footnote[1]{aron@ens.fr},
Manas Kulkarni\textsuperscript{2}
\end{center}

\begin{center}
{\bf 1} 
Laboratoire de Physique, \'Ecole Normale Sup\'erieure, CNRS, \\
Universit\'e PSL, Sorbonne Universit\'e, Universit\'e de Paris, 75005 Paris, France
\\ \medskip
{\bf 2} International Centre for Theoretical Sciences, \\ Tata Institute of Fundamental Research, 560089 Bangalore, India
\end{center}

\section*{Abstract}
{\bf
Many-body non-equilibrium steady states can be described by a Landau-Ginzburg theory if one allows non-analytic terms in the potential. We substantiate this claim by working out the case of the Ising magnet in contact with a thermal bath and undergoing stochastic reheating: It is reset to a paramagnet at random times.
By a combination of stochastic field theory and Monte Carlo simulations, we unveil how the usual $\varphi^4$ potential is deformed by non-analytic operators of intrinsic non-equilibrium nature. We demonstrate their infrared relevance at low temperatures by a renormalization-group analysis of the non-equilibrium steady state.
The equilibrium ferromagnetic fixed point is thus destabilized by stochastic reheating, and we identify the new non-equilibrium fixed point.
}

\section{Introduction} \label{sec:intro}

The Landau-Ginzburg theory is a monumental cornerstone of modern physics that unifies the various equilibrium phase transitions of matter\footnote{Exception being topological phase transitions.} in a common framework.
It relies on the effective description of many-body systems in terms of a local order-parameter field, say $\varphi(x)$. The probability distribution at equilibrium (EQ) is given by $P_{\rm EQ}[\varphi] \sim \exp(-\mathcal{F}_{\rm EQ}[\varphi])$, where the free-energy functional $\mathcal{F}_{\rm EQ}[\varphi]$ is built on simple principles: locality, symmetry, stability, and analyticity. Subsequently, $\mathcal{F}_{\rm EQ}[\varphi]$ can be fed to a renormalization group (RG) analysis in order to access the universal infrared features of the many-body system.

Away from thermal equilibrium, non-equilibrium phase transitions can be studied in the framework of non-equilibrium extensions to the dynamical field theories classified by Hohenberg and Halperin~\cite{Hohenberg}. The theories that have perhaps received the most attention are those for which the dynamics conserve a global quantity such as the particle number, \textit{i.e.} whose field-theoretic description revolves around the Model B of Hohenberg and Halperin. Examples include driven-diffusive systems~\cite{Zia, Tauber} such as the driven lattice gas~\cite{Katz,Dickman}, but also active matter systems~\cite{Cates} displaying, \textit{e.g.}, phase separation or pattern formation~\cite{Tailleur1, Poon}.
Another class of non-equilibrium systems is the so-called driven-dissipative systems, with no conserved quantity. This includes growth processes, such as the directed percolation~\cite{Kinzel,Hinrichsen} or the Kardar-Parisi-Zhang problems~\cite{KPZ,Lassig}. 

While the quest for a unified principle-based approach to all those non-equilibrium theories still seems out of reach, a more reasonable ambition is to restrict the focus to non-equilibrium steady states (NESS), \textit{i.e.} time-translational invariant non-thermal states described by a stationary probability measure $P_{\rm NESS}[\varphi]$.
Recent developments have led us to propose the following claim.
\begin{theorem}
A Landau-Ginzburg theory can be developed for NESS at the cost of abandoning the principle of analyticity of the Landau-Ginzburg functional $\mathcal{F}_{\rm NESS}[\varphi] \equiv - \log P_{\rm NESS}[\varphi]$
\end{theorem} 

This idea emerged out of a heuristic observation in the context of a driven-dissipative quantum antiferromagnet with a $\mathbb{Z}_2$-symmetric staggered order parameter $\varphi$~\cite{Jong}.

The idea was then sharpened in Ref.~\cite{Claudio} where the claim was formulated in the framework of the Ising magnet undergoing driven-dissipative dynamics. The explicit computations were made possible by a mean-field approximation: The magnetization $\varphi$ was assumed to be homogeneous throughout the magnet.
The static steady state $P_{\rm NESS}(\varphi) \sim \exp(-\mathcal{V}_{\rm NESS}(\varphi))$ was computed in a couple of simple concrete examples, such as a two-bath Ising model. They unveiled non-analytic Landau potentials of the form
\begin{align}
\mathcal{V}_{\rm NESS}(\varphi) = \underbrace{a_2 \varphi^2 + a_4 \varphi^4}_{\rm analytic} + \underbrace{c_\alpha |\varphi|^{2+\alpha}}_{\rm non-analtyic} + \ldots
\end{align}
where the exponent $\alpha$, with $0 < \alpha < 2$, depends on the low-energy features of the environment and can be non-integer valued. The coefficients $a_2$, $a_4$, and $c_\alpha$ are smooth functions of the external parameters, and $c_\alpha$ vanishes at equilibrium. The non-analytic terms are therefore of intrinsic non-equilibrium nature\footnote{Conversely, the absence of such non-analytic terms does not imply equilibrium.}.

The results of Ref.~\cite{Claudio} definitely introduced an important conceptual milestone to the construction of a field theory for non-equilibrium steady states, specifically the loss of analyticity of the Landau potential. This heralds a possible extension of the concept of universality class to non-equilibrium steady states: Critical exponents  are determined not only by the dimensions of the system and the symmetries of its order parameter, but also by the low-energy features of its environment.

However, questions remain to ascertain that these findings are not artifacts of the mean-field approximation that was used, but are indeed robust non-equilibrium field-theoretic hallmarks present in finite dimensions:
\begin{enumerate}
\item Is there a finite upper critical dimension $d_{\rm uc}$ above which fluctuations are irrelevant to the non-equilibrium steady state and the mean-field approximation is exact?
\item Do the non-analytic terms of the non-equilibrium Landau potential survive in the infrared (IR)? Or are they washed out when the order parameter is coarsegrained on larger and larger scales?
\item Do the non-analytic terms survive fluctuations below the upper critical dimension?
\end{enumerate}

In this paper, we address these general questions in the specific framework of the driven-dissipative dynamics of the Ising model undergoing stochastic reheating. We introduce the non-equilibrium model and its field-theoretic description in Sec.~\ref{sec:Ising}. In Sec.~\ref{sec:above}, we present analytical arguments that give positive and unambiguous answers to the first two questions above.
Importantly, these positive answers validate the claim stated above beyond the scope of the specific non-equilibrium dynamics studied here.
We close in Sec.~\ref{sec:discussion} by presenting solid numerical arguments in favor of a positive answer to the third question, and by discussing possible routes to complete the field-theoretic picture initiated in this work.

\medskip

Let us summarize the key results:
\begin{itemize}
\item the field theory for non-equilibrium steady states in finite dimensions is found to be a deformation of the equilibrium $\varphi^4$ theory by non-analytic operators
\item the non-analytic operators are shown to be IR relevant, or IR irrelevant, depending on the bath temperature
\item $d_{\rm uc}$ is conjectured to be 4, \textit{i.e.} the same as the equilibrium upper critical dimension
\item above $d=4$, the Landau potentials at the IR fixed points are computed exactly
\item supporting evidence from Monte Carlo numerics is provided in both $d=4$ and $d=2$.
\end{itemize}
 
\section{Kinetic Ising model with stochastic reheating}
\label{sec:Ising}

We address the questions listed above in a simple framework: a driven-dissipative many-body system with no conserved quantity and whose non-equilibrium steady state is homogeneous and isotropic.
Specifically, we consider the infamous $\mathbb{Z}_2$-symmetric Ising magnet whose relaxation dynamics are generated by the coupling to an equilibrium bath at temperature $T$ --inducing thermal spin flips-- and whose non-equilibrium drive is realized by a \emph{stochastic reheating} protocol: The magnet is randomly reheated, with rate $r$, to an infinite-temperature paramagnetic state. Between two reheating events, the dynamics are the ones of a quench from infinite temperature to the bath temperature $T$.

The choice of stochastic reheating as the non-equilibrium drive is guided by the expectation that the mean-field description of our many-body problem reduces to a diffusive single particle undergoing so-called \emph{stochastic resetting}: It is reset to a given position at random times with rate $r$~\cite{EM1}.
The generation of non-equilibrium states by stochastic resetting is a simple alternative compared to traditional setups where multiple reservoirs are attached to the system. Over the past decade, a body of exact results has already been obtained in different implementations ranging from space- or time-dependent resetting rate~\cite{pke}, resetting to a random position~\cite{EM2}, generalizations to higher dimensions~\cite{EMhigherD} and extended systems in the Kardar-Parisi-Zhang universality class~\cite{gmskpz}.
The use of stochastic resetting as a non-equilibrium drive to many-body systems was pioneered very recently in Ref.~\cite{MMS20} where, contrary to our case, the Ising model is reset to an ordered ferromagnetic state, yielding much different dynamics and ensuing non-equilibrium steady states.

\subsection{Instability of the ferromagnetic fixed point}
\paragraph{Equilibrium}
The absence of stochastic reheating, when $r=0$, corresponds to thermal equilibrium. The Ising model undergoes Glauber dynamics governed by the detailed balance at the temperature $T$ of the bath. This is often referred to as the kinetic Ising model. The corresponding field-theoretic description is given by the well-known $O(n=1)$-symmetric $\varphi^4$ theory~\cite{Kardar}, where the real field $\varphi(x)$ is a coarsegrained measure of the magnetization, say on cells of linear size $l$~\cite{binder81}.
The scale $l$ plays the role of the ultraviolet (UV) cutoff of the field theory.
Cooling down the bath temperature across the critical temperature $T_{\rm c}$, the phase transition from the paramagnetic phase to the ferromagnetic phase is described by the spontaneous symmetry breaking of the $\mathbb{Z}_2$ symmetry when the shape of the Landau potential changes from a parabola to a double-well Mexican hat.

\begin{figure}
\center
\includegraphics[scale=0.7]{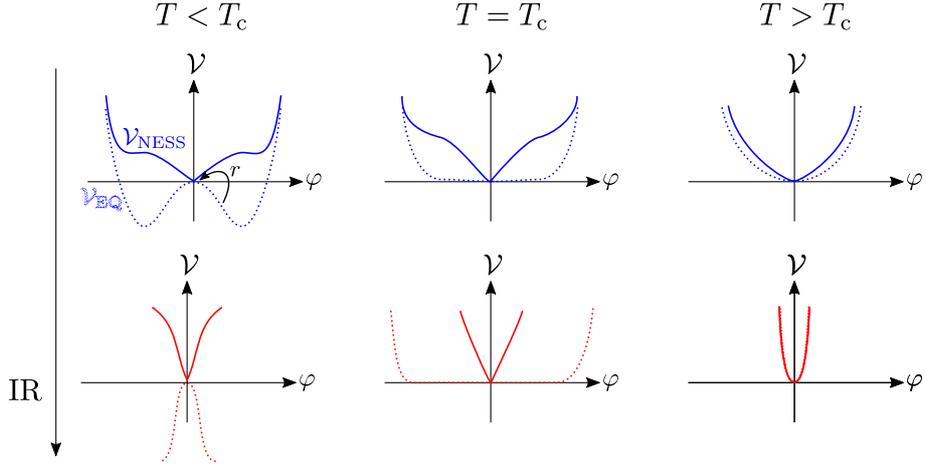}
\caption{
Schematic of the renormalization of the non-equilibrium steady-state Landau potential $\mathcal{V}_{\rm NESS}(\varphi)$ of the Ising magnet undergoing stochastic reheating with rate $r$ and in $d \geq 4$. The red color is used for quantities in the infrared (IR) regime.
At bath temperatures $T > T_\rmc$, the infrared fixed point is simply the equilibrium fixed point of Eq.~(\ref{eq:VNESS_HT}). For $T \leq T_\rmc$, a cusp develops at $\varphi=0$ yielding a new non-equilibrium fixed point. The corresponding non-analytic fixed-point potentials are computed explicitly in Eqs.~(\ref{eq:VNESS_LT}) and (\ref{eq:VNESS_Tc}). The equilibrium scenario $(r=0)$ is recalled with dotted lines.
\label{fig:sketch}
}
\end{figure}

\paragraph{Non-equilibrium steady state} \label{sec:simple}
The case of interest, when the reheating rate $r>0$, corresponds to non-equilibrium dynamics. After a transient regime which depends on the initial state preparation, the many-body dynamics are expected to reach a non-equilibrium steady state (NESS) described by a static probability distribution $P_{\rm NESS}[\varphi]$.

Importantly, we can readily argue that a finite reheating rate, $r>0$, is a singular non-equilibrium perturbation in the sense that it changes drastically the physics of the ordered phase that is found at equilibrium ($r=0$).
Indeed, the stochastic reheating protocol effectively restores the $\mathbb{Z}_2$ symmetry that is spontaneously broken in equilibrium at low temperatures.
This simple observation has important consequences:
\begin{itemize}
\item[(i)] at fixed $r>0$, the average local magnetization vanishes identically in the steady state, irrespective of the bath temperature: $\langle \varphi(x) \rangle=0$ where the average is taken with respect to $P_{\rm NESS}$.
However, the vanishing of the order parameter at all temperatures must not hide the presence of two distinct non-equilibrium phases: 
\begin{itemize}
\item[(a)] a paramagnetic phase at high temperatures, $T> T_\rmc$, where the system is always and everywhere paramagnetic.
There, the stochastic reheating protocol is expected to be mostly inconsequential since those long wavelength modes with relaxation timescales longer than the reheating timescale set by $1/r$ mostly live at infinite temperature, while the modes with timescales shorter than $1/r$ have the time to thermalize to the temperature $T$,
\item[(b)] an ordering phase at low temperatures, $T < T_\rmc$, where after each reheating event the system undergoes the much-studied coarsening dynamics of a ferromagnet after a quench from infinite temperature and across the phase transition~\cite{Bray, Gambassi, Alberto}: ferromagnetic domains of opposite magnetization compete and grow as $\xi(t)\sim\sqrt{t}$. The stochastic reheating halts this growth, yielding a typical correlation length $\xi_{\rm max} \sim 1/\sqrt{r}$.
\end{itemize}
These phases are not distinguishable by the vanishing global order parameter, but they can be captured, \textit{e.g.}, by the fluctuations of local order parameter.
\item[(ii)] at bath temperatures $T < T_\rmc$, there is a discontinuous phase transition when turning on $r$ between the equilibrium ferromagnet (with $\langle \varphi(x) \rangle \neq 0$) and the non-equilibrium steady state (with $\langle \varphi(x) \rangle=0$),
\item[(iii)] at finite $r$ and at scales larger than the correlation length, $l \gtrsim \xi_{\rm max}$, the corresponding non-equilibrium Landau-Ginzburg functional can only have a single global minimum at $\varphi = 0$.
\end{itemize}
From an RG viewpoint, this means that stochastic reheating is a relevant perturbation that destabilizes the equilibrium low-temperature (Gaussian or Wilson-Fisher) fixed point, irrespective of the dimension.

 \subsection{Correspondence with quench dynamics}
In the presence of stochastic reheating with rate $r$, the dynamics of the field from time $t$ to $t+\rmd t$ are governed by the following Poisson process: 
\begin{itemize}[topsep=4pt,itemsep=3pt,partopsep=4pt, parsep=4pt]
 \item[-] with probability $1- r \, \rmd t$, the field $\varphi(x,t+\rmd t)$ is set by the usual stochastic dynamics in the \emph{absence} of reheating,
 \item[-] with probability $r \, \rmd t$, the field is set to  $\varphi(x,t+\rmd t) = 0 $ for all $x$.
\end{itemize}
The associated Fokker-Panck equation governing the evolution of the probability distribution of the magnetization field, $P([\varphi];t)$, reads~\cite{EM1}
\begin{align} \label{eq:FP_func}
\partial_t P([\varphi];t) &= - \mathcal{L}_0[\varphi]  \cdot P([\varphi];t) - r \,  P([\varphi];t) +r \,\delta[\varphi] \,,
\end{align}
where $\mathcal{L}_0[\varphi]$ is the (linear) Fokker-Planck operator generating the time evolution of the probability distribution in the \emph{absence} of reheating (\textit{i.e.} $r=0$), to be discussed below in Eq.~(\ref{eq:FPO}).

\label{sec:correspondence}
\paragraph{Renewal formula}
The task of computing the stationary measure $P_{\rm NESS}[\varphi] \equiv \lim\limits_{t\to\infty} P([\varphi];t)$ of the stochastic reheating problem in Eq.~(\ref{eq:FP_func}) can be much simplified by realizing that it is of the form
\begin{align} \label{eq:renewal}
P_{\rm NESS}[\varphi] = r \int_0^{\infty} \Ud{t} \rme^{-r t} P_0([\varphi]; t)\,,
\end{align}
where $P_0([\varphi];t)$ is the time-dependent probability distribution after a quench from infinite temperature and in the \emph{absence} of reheating.
Indeed, inserting Eq.~(\ref{eq:renewal}) into Eq.~(\ref{eq:FP_func}), one can check that $P_0([\varphi];t)$  is the solution of the Fokker-Planck equation
\begin{align} \label{eq:FP0}
\partial_t P_0([\varphi];t) &=  - \mathcal{L}_0[\varphi]  \cdot P_0([\varphi];t)  \,,
\end{align}
with the initial condition $P_0([\varphi], t=0) = \delta[\varphi]$.
These purely dissipative dynamics will be briefly described below. 

Equation~(\ref{eq:renewal}) is referred to as the renewal formula~\cite{EM1, EMS} and it has important implications. 
It connects the statics of a fully non-equilibrium problem to the relaxation dynamics of the field in contact with an equilibrium bath.
As we shall see in Sec.~\ref{sec:above}, this relation is instrumental to performing an RG analysis in the non-equilibrium steady state (NESS) using the knowledge of what is known about the relaxation dynamics of magnetization.
In particular, it tells us that field fluctuations can be neglected in the NESS whenever it is legitimate to neglect them in the relaxation dynamics; we therefore expect the upper critical dimension of the stochastic reheating problem to be $d_{\rm uc}= 4$. 
From a numerical perspective, the renewal formula also implies that $P_{\rm NESS}[\varphi]$ can be simply reconstructed from the data of a quench up to finite times on the order of a few $1/r$. It is worth noticing that although the integrand in Eq.~(\ref{eq:renewal}) may be analytic in the field, the integral itself can be non-analytic.

\paragraph{Model A relaxation}
The dynamics of magnetization after a quench from infinite temperature are described by the initial condition $\varphi(x,t=0) = 0$ and by the so-called Model A dynamics~\cite{Hohenberg, Tauber} given by the stochastic equation
\begin{align} \label{eq:modelA}
\eta \partial_t \varphi(x,t) = - \frac{\delta \mathcal{F}_{\rm EQ} [\varphi] }{\delta \varphi(x,t)} + \xi(x,t)\,,
\end{align}
where $\eta > 0 $ is a dimensionless ``friction'' parameter, and $\xi$ is a Gaussian white noise with $\langle \xi(x,t)\rangle = 0$ and $\langle \xi(x,t) \xi(x',t') \rangle = 2 \eta \, \delta(x-x')\delta(t-t')$. 
Equivalently, the Model A dynamics can be characterized by their associated Fokker-Planck equation Eq.~(\ref{eq:FP0}) with the Fokker-Planck operator
\begin{align} \label{eq:FPO}
\mathcal{L}_0[\varphi] \; \bullet  = - \frac{1}{\eta} \int \rmd^d x \; \frac{\delta}{\delta \varphi(x)} \left[ \frac{\delta \mathcal{F}_{\rm EQ}[\varphi]}{\delta \varphi(x)} \; \bullet \  + \frac{\delta}{\delta \varphi(x)} \; \bullet \ \ \right] \,.
\end{align}
$\mathcal{F}_{\rm EQ}[\varphi]$ is the equilibrium Landau-Ginzburg free-energy functional ensuring that the stationary measure of the stochastic process is the equilibrium measure: $\lim\limits_{t\to\infty}P_0([\varphi];t) = P_{\rm EQ}[\varphi] \sim \exp(-\mathcal{F}_{\rm EQ}[\varphi])$. Following the set of principles of the Landau-Ginzburg theory, the expression for $\mathcal{F}_{\rm EQ}[\varphi]$ is local (it involves a local free-energy density, function of $\varphi$ and $\nabla \varphi$), $O(n=1)$ symmetric, stable (it involves a confining potential), and analytic in $\varphi$ and $\nabla \varphi$. This yields the usual $\varphi^4$ theory, also called the Ginzburg-Landau-Wilson model, reading
\begin{align} \label{eq:F_eq}
\mathcal{F}_{\rm EQ}[\varphi] = \int \rmd^d x \;
\left( 
\frac12 \mu \varphi^2 + \frac{1}{4} \lambda \varphi^4 + \ldots
+ \frac12 ({\nabla} \varphi)^2 + \ldots \right)\,.
\end{align}
The various static parameters $\mu$, $\lambda$, etc., as well as the dynamic parameter $\eta$ depend on the coarsegraining scale $l$. The ``mass'' $\mu$ corresponds to the distance to criticality, \textit{i.e.} $\mu \sim T-T_{\rm c}$, which diverges in the infrared away from criticality: $\mu(T \gtrless T_\rmc) \stackrel{\rm IR}{\longrightarrow} \pm \infty$.

\subsection{Non-equilibrium steady-state Landau-Ginzburg functional}
\label{sec:anticipation}
In the following sections, we shall argue that the non-equilibrium steady-state Landau-Ginzburg functional, $\mathcal{F}_{\rm NESS}[\varphi] \equiv -\log P_{\rm NESS}[\varphi]$, is of the form
\begin{align} \label{eq:F_NESS} 
\mathcal{F}_{\rm NESS}[\varphi] = \int \rmd^d x \;
\left( c_1 |\varphi| + a_2 \varphi^2 + c_3 |\varphi|^3 + a_4 \varphi^4 + \ldots + \frac12 (\nabla \varphi)^2 + \ldots \right)\,.
\end{align}
This form follows the usual Landau-Ginzburg set of principles recalled above in Sec.~\ref{sec:correspondence} except for the principle of analyticity of the potential that has been abandoned. Indeed, the terms in $|\varphi|$, $|\varphi|^3$, etc., are non-analytic at $\varphi = 0$. They are of intrinsic non-equilibrium nature and their coefficient, $c_1$, $c_3$, etc., must vanish at thermal equilibrium (when $r=0$) in order to recover the equilibrium Landau-Ginzburg-Wilson theory. The possible presence, below the upper critical dimension, of non-conventional non-analytic terms in the gradient expansion, will be discussed in Sec.~\ref{sec:discussion}

Anticipating the results that follow, we sketch the renormalization of the Landau potential in Fig.~\ref{fig:sketch}. 
While above the critical temperature, the infrared fixed point is the equilibrium (paramagnetic) fixed point, the non-analytic terms are relevant below the critical temperature. We shall compute explicitly the corresponding non-analytic fixed-point potential above the upper critical dimension.

\section{Above the upper critical dimension} \label{sec:above}

In this section, we construct the non-equilibrium field theory of the Ising model undergoing stochastic reheating dynamics in dimensions above the equilibrium upper critical dimension, $d_{\rm uc}=4$.
In those dimensions, the spatial fluctuations of the equilibrium problem around its mean-field solution can be neglected. Thus, making use of the renewal formula (\ref{eq:renewal}), solving our non-equilibrium many-body problem reduces to solving a single-particle problem: a diffusive particle undergoing the Langevin dynamics
\begin{align} \label{eq:Langevin}
\eta \partial_t \varphi(t) = - \partial_\varphi \mathcal{V}_{\rm EQ}(\varphi) + \xi(t)\,,
\end{align}
with the initial condition $\varphi(0) = 0$, and where $\eta > 0$ is a ``friction'' parameter, $\xi$ is a Gaussian white noise with correlator $\langle \xi(t) \xi(t') \rangle = 2 \eta \, \delta(t-t')$, and $\mathcal{V}_{\rm EQ}(\varphi) = \frac{\mu}{2} \varphi^2 + \frac{\lambda}{4} \varphi^4 + \ldots $ is the equilibrium potential of a coarsegraining cell. Here, we redefined the coefficients $\eta$, $\mu$, $\lambda$, etc., that were introduced in Eqs.~(\ref{eq:modelA}) and~(\ref{eq:F_eq}) to include the volume of the coarsegraining cell.
Furthermore, at $d_{\rm uc} =4$ and above, all the infrared equilibrium fixed points of the RG are Gaussian, that is
\begin{align} \label{eq:quadratic_V_EQ}
\mathcal{V}_{\rm EQ}(\varPhi) \stackrel{\rm IR}{\longrightarrow} \mathcal{V}^*_{\rm EQ}(\varPhi) =
\left\{ 
\begin{array}{ll}
 + \frac{1}{2} \varPhi^2 & \mbox{ for } T > T_\rmc \mbox{ (paramagnet)} \\
0 & \mbox{ at }T= T_\rmc \mbox{ (critical)}\\
- \frac{1}{2} \varPhi^2 + 0^+ \varPhi^4 & \mbox{ for } T < T_\rmc \mbox{ (ferromagnet)}
\end{array} 
\right. \,,
\end{align}
where we introduced the rescaled fields $\varPhi \equiv \sqrt{\eta} \, \varphi$ for the critical case $T = T_\rmc$ and $\varPhi \equiv \sqrt{|\mu|} \, \varphi$ for the non-critical cases $T \neq T_\rmc$. In the latter cases, it will be also useful to introduce the reduced reheating rate $R \equiv r \, \eta / |\mu|$ which flows to 0 in the infrared limit.
Owing to the linearity of the resulting Langevin equation, the infrared limit of the non-equilibrium steady state, $P^*_{\rm NESS}(\varPhi)$, can be computed exactly.
Equivalently, $P^*_{\rm NESS}(\varPhi)$ can be directly computed by solving the stationary Fokker-Planck equation in the presence of stochastic reheating, namely
\begin{align}
0 = \partial_\varPhi \left[ \partial_\varPhi \mathcal{V}^*_{\rm EQ}(\varPhi) \,  P^*_{\rm NESS}(\varPhi) + \partial_\varPhi P^*_{\rm NESS}(\varPhi) \right] - R \, P^*_{\rm NESS}(\varPhi) + R \, \delta(\varPhi)  \,.
\end{align}
We refer the reader to Ref.~\cite{Arnab} for details of such computations.
As it was already predicted in Sec.~\ref{sec:simple} using general arguments, the form of the resulting effective Landau potential $\mathcal{V}^*_{\rm NESS}(\varPhi)$ will markedly depend on whether the bath temperature $T$ is above or below $T_{\rm c}$.
 
It is noteworthy that, even before reaching the IR fixed point, the equilibrium potential $\mathcal{V}_{\rm EQ}$ may still be truncated to the quadratic $\mathcal{V}^*_{\rm EQ}$ in Eq.~(\ref{eq:quadratic_V_EQ}) when the reheating rate is large enough, \textit{i.e.} when the diffusive particle does not have time to experience the effect of the non-linear terms before being reset to $\varphi =0$. We may roughly estimate the minimum reheating rate above which one may neglect the $\varphi^4$ term in the potential as $R_{\rm min} \sim1/ \log \left( {\mu^2}/{\lambda} \right)$. 

\begin{figure} 
\center
\input{FxPt.tex}
\hspace{-2em}
\input{FxPt_Tc.tex}
\caption{
Infrared non-equilibrium fixed-point potential $\mathcal{V}^*_{\rm NESS}$. (a) Below $T_\rmc$ (solid line), the non-analytic potential is given in Eq.~(\ref{eq:VNESS_LT}). Above $T_\rmc$ (dotted lines), the paramagnetic equilibrium fixed point is recovered, see Eq.~(\ref{eq:VNESS_HT}).
(b) At $T_\rmc$, the non-analytic potential depends on the reheating rate $r$.
\label{fig:VNESS}}
\end{figure}
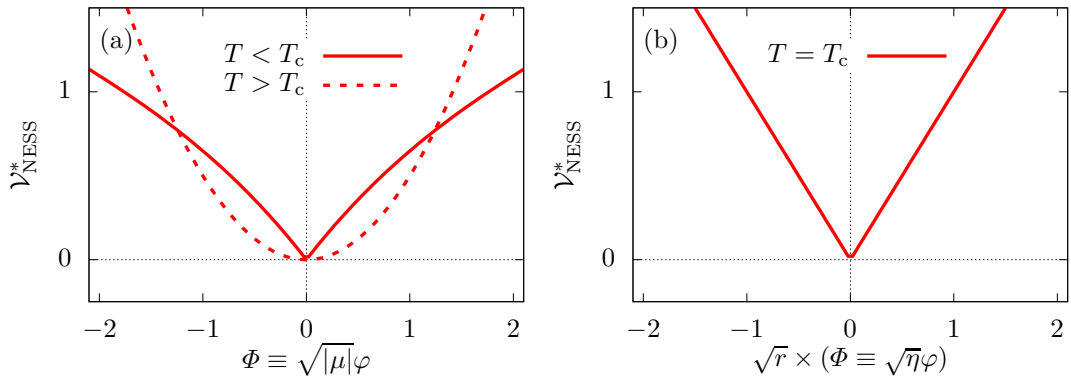

\subsection{High-temperature phase}
For bath temperatures above the critical temperature, $T >T_{\rm c}$, and at scales for which the equilibrium system would be in the vicinity of its paramagnetic fixed point, we find the non-equilibrium steady state potential
\begin{align}
\mathcal{V}_{\rm NESS}(\varPhi) &= \frac12 {\varPhi^2} 
- \log \left[ H\left(-R, \frac{|\varPhi|}{\sqrt{2}} \right)
 \right] + \mbox{const} \label{eq:V_NESS_above}\\
 & = \mbox{const} + c_1 |\varPhi| + a_2 \varPhi^2 + c_3 |\varPhi|^3 + \ldots\,, \label{eq:V_NESS_above2}
\end{align}
with $\varPhi\equiv \sqrt{\mu} \varphi$ and where $H(n,x)$ is the Hermite polynomial of degree $n$. The coefficients $c_1 \equiv \sqrt{2} \frac{\Gamma\left(\frac{1+R}{2}\right)}{\Gamma(R/2)} \geq 0 $, $1/2 \geq a_2 \equiv \frac{1 - R}{2}+ \frac{c_1^2}{2} > 1/4$, and $c_3 \equiv \frac{c_1}{3} \left(c_1^2 + 1/2 -R\right) \geq 0$.
The terms in $|\varPhi|$, $|\varPhi|^3$, etc., are non-analytic around $\varPhi = 0$. They are intrinsically non-equilibrium in nature and one can check that their coefficients $c_1$, $c_3$, etc., vanish in the equilibrium limit (\textit{i.e.}, in the limit $r \to 0$). The expression in Eq.~(\ref{eq:V_NESS_above2}) justifies the form of the non-equilibrium steady-state Landau-Ginzburg functional $\mathcal{F}_{\rm NESS}[\varphi]$ that was anticipated in Sec.~\ref{sec:anticipation}.

Finally, we can identify the IR fixed point of the non-equilibrium steady state by sending $\mu\to\infty$ in Eq.~(\ref{eq:V_NESS_above}). Using the property $H(0,x)=1$, this simply gives back the equilibrium potential
\begin{align}
\mathcal{V}_{\rm NESS}(\varPhi) \stackrel{\rm IR}{\longrightarrow} \mathcal{V}^*_{\rm EQ}(\varPhi)= \frac{1}{2} \varPhi^2 \label{eq:VNESS_HT}
\end{align}
showing that the non-analytic terms in the potential are IR irrelevant above the critical temperature.
We can therefore conclude that the paramagnetic equilibrium fixed point is robust against stochastic reheating.

\subsection{Low-temperature phase}
Let us now consider bath temperatures below the critical temperature, $T < T_{\rm c}$. At scales for which the equilibrium system would be in the vicinity of its ferromagnetic fixed point, we find
\begin{align} \label{eq:V_NESS_below}
\mathcal{V}_{\rm NESS}(\varPhi) &= 
-\log \left[ H\left(-1-R, \frac{|\varPhi|}{\sqrt{2}} \right) \right] + \mbox{const} \\
& = \mbox{const}+ c_1 |\varPhi| + a_2 \varPhi^2 + c_3 |\varPhi|^3 + \ldots\,,
\end{align}
with $\varPhi \equiv \sqrt{-\mu} \varphi$ and the coefficients $c_1 \equiv \sqrt{2} \, \frac{ \Gamma \left(1+ \frac{R}{2}\right)}{\Gamma \left(\frac{1+R}{2}\right)} \geq 0$, $a_2 \equiv \frac{c_1^2}{2} - \frac{1+R}{2} < 0 $, and $c_3 \equiv \frac{c_1}{3} \left( c_1^2 - 1/2 - R \right) >0$.
Similarly to the high-temperature case, the effective Landau potential displays non-analytic features of intrinsic non-equilibrium nature around $\varPhi = 0$, justifying the form of $\mathcal{F}_{\rm NESS}[\varphi]$ proposed in Eq.~(\ref{eq:F_NESS}).

The IR fixed point of the non-equilibrium steady state can be accessed by sending $\mu\to-\infty$ in Eq.~(\ref{eq:V_NESS_below}). This yields the non-equilibrium fixed-point potential
\begin{align} \label{eq:VNESS_LT}
\mathcal{V}_{\rm NESS}(\varPhi) \stackrel{\rm IR}{\longrightarrow} 
\mathcal{V}^*_{\rm NESS}(\varPhi) &= -\log \left[ H\left(-1, \frac{|\varPhi|}{\sqrt{2}} \right) \right] + \mbox{const} \\
 & = \mbox{const} + \sqrt{\frac{2}{\pi}} |\varPhi|
- (1/2-1/\pi) \varPhi^2
+ \frac{4/\pi-1 }{3 \sqrt{2 \pi} } |\varPhi|^3 + \ldots\,.
\end{align}
This non-analytic infrared fixed-point potential is one of the main results of this paper. It is plotted in Fig.~\ref{fig:VNESS}. 
It demonstrates that, contrary to the high-temperature phase discussed above, non-analyticities originally present in the microscopic Landau potential $\mathcal{V}_{\rm NESS}(\varPhi)$ survive in the infrared. In particular, the term in $|\varPhi|$ dominates $\mathcal{V}^*_{\rm NESS}(\varPhi)$ around the global minimum $\varPhi=0$, yielding a distinctive cusp. Notably, $\mathcal{V}^*_{\rm NESS}(\varPhi)$ does not depend on the reheating rate any longer.

\subsection{Critical-temperature state} 
When the bath temperature is exactly at $T=T_{\rm c}$, following the single-particle computation in Ref.~\cite{EM1}, we find the infrared non-equilibrium fixed-point potential
\begin{align} \label{eq:VNESS_Tc}
\mathcal{V}_{\rm NESS}^*(\varPhi) = \sqrt{r} \, |\varPhi| + \mbox{const}\,,
\end{align}
with $\varPhi \equiv \sqrt{\eta} \varphi$.
Contrary to the high-temperature and low-temperature cases above, this infared fixed-point potential still explicitly depends on the reheating rate $r$. 
This was expected since the reheating protocol introduces a finite correlation length $\xi_{\rm max} \sim 1/\sqrt{r}$ which breaks the scale invariance found at equilibrium.
The equilibrium fixed-point potential is recovered in the limit $r \to 0$.

\subsection{Supporting Monte Carlo results}

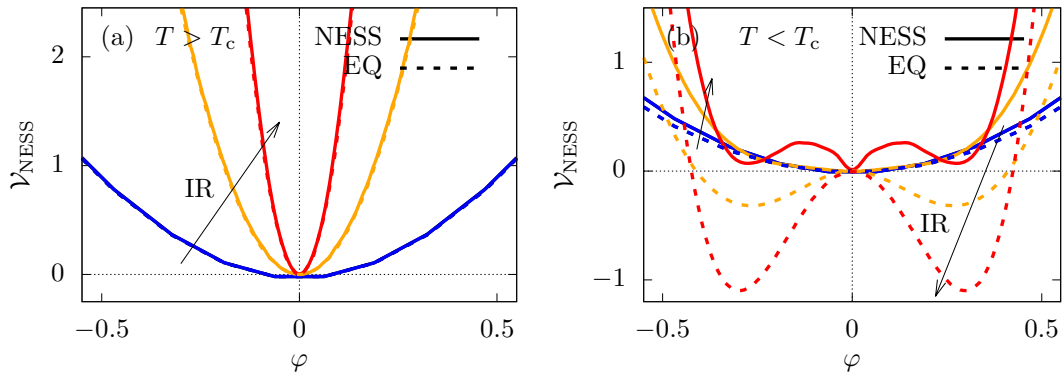
\begin{figure}
\center
\input{RG_HT.tex}
\hspace{-2em}
\input{RG_LT.tex}
\caption{\label{fig:MC-NESS}
The $d=4$ Monte Carlo solution of the non-equilibrium steady-state potential $\mathcal{V}_{\rm NESS}(\varphi)$ for increasing coarsegraining lengths: $l = 2,\, 4,\, 8$ ($L=8$, $r=0.005$).
(a) Above $T_\rmc$ ($T = 1.1 \, T_\rmc$), $\mathcal{V}_{\rm NESS}$ (solid lines) does not differ from the equilibrium potential $\mathcal{V}_{\rm EQ}$ (dotted lines).
(b) Below $T_\rmc$ ($T = 0.99 \, T_\rmc$), $\mathcal{V}_{\rm NESS}$ deviates markedly from the equilibrium Mexican hat. A cusp develops at the new minimum $\varphi=0$ which becomes the global minimum in the infrared (IR).
}
\end{figure}

The theory developed so-far relied on a few assumptions. The main assumption was the validity of the mean-field approximation above the upper critical dimension. The latter was conjectured to be identical to the equilibrium upper critical dimension, \textit{i.e.} $d_{\rm uc} = 4$.

To support all the steps undertaken so far, we numerically solve for the many-body non-equilibrium steady state, constructing the field theory from a numerical standpoint, void of any assumption.
All the numerical computations that we present below use well-established methods and required moderate computing power to be conclusive. This leaves room for future interesting in-depth analysis of the RG flow via high performance computing.

\paragraph{Details of the numerics}
We consider the Ising model
\begin{align}
H = - \sum_{\langle i j\rangle} S_i S_j\,,
\end{align}
where the $N =L^d$ classical spins $S_{i} = \pm 1 $ are located on a $d$-dimensional cubic lattice of linear length $L$ with periodic boundary conditions and interact ferromagnetically between nearest neighbors.
The values of $L$ we use are indicated in the captions of the different figures.
Here, to directly compare with the theory developed in Sec.~\ref{sec:above}, we work in four dimensions: $d=4$. In Sec.~\ref{sec:discussion}, we shall also present numerical results below the upper critical dimension, in $d=2$.
Due to the lack of a well-developed finite-size scaling theory for non-equilibrium steady states, we cannot afford to correct for finite-size effects. We therefore stay slightly away from the critical temperature $T_c(d=4) \approx 6.68$ by working with $T = 0.99 \, T_\rmc$ and $T = 1.1 \, T_\rmc$.

The dynamics due to the local thermal baths are generated by the usual Monte Carlo Metropolis algorithm: A local spin flip $S_i \to - S_i$ is accepted with probability $\mathrm{min}(1, \exp({-2 S_i h_i/ T} ))$, where $h_i$ is the local instantaneous Weiss field.
The reheating dynamics are taken into account via the renewal formula~Eq.~(\ref{eq:renewal}). This amounts to following the dynamics of the local magnetization
\begin{align}
\phi(x,t) = \frac{1}{l^d} \sum_{i \in v_l(x)} S_i(t) \,,
\end{align}
where the microscopic spin degrees of freedom are averaged on cubic coarsegraining cells $v_l(x)$ of linear size $l$ around $x$, 
after a quench from an infinite-temperature disordered initial state, $S_i(t=0) = \pm 1$ with probability $1/2$. Here, the time $t$ is counted in units of Monte Carlo steps defined as a sequence of $N$ attempted spin flips.
Rather than working with the magnetization which is bounded, $\phi \in [-1, 1]$, we perform a field redefinition and introduce the field
\begin{align}
\varphi & \equiv {\rm artanh}(\phi) \in \mathbb{R} \,.
\end{align}
The non-equilibrium Landau potential $\mathcal{V}_{\rm NESS}(\varphi)$ is computed from the Monte Carlo data by first measuring the time-dependent probability distribution of the local magnetization after the quench, $P_{\rm 0}(\varphi ; t)$.
In practice, we improve the statistics by averaging $P_{\rm 0}(\varphi ; t)$ over many realizations of the quench dynamics. Then, the renewal formula in Eq.~(\ref{eq:renewal}) is used to compute the non-equilibrium steady-state distribution $P_{\rm NESS}(\varphi) $, and finally $\mathcal{V}_{\rm NESS}(\varphi) \equiv - \log P_{\rm NESS}(\varphi)$.

Note that, in order to reconstruct the full Landau-Ginzburg functional $\mathcal{F}_{\rm NESS}[\varphi]$, one should in principle extract the joint probability distribution of the field and its gradients, $P_{\rm 0}(\varphi, \nabla \varphi ; t)$. Here, operating above the upper critical dimension, we discard the information on the gradients and we concentrate on comparing $\mathcal{V}_{\rm NESS}(\varphi)$ with the theory developed in Sec.~\ref{sec:above}. We shall return to the possibility of non-conventional gradient terms present in the Landau-Ginzburg functional below the upper critical dimension in the discussion of Sec.~\ref{sec:discussion}.

\paragraph{Agreement with theory}
In Fig.~\ref{fig:MC-NESS}, we plot the non-equilibrium steady-state potential $\mathcal{V}_{\rm NESS}(\varphi)$ for different values of the coarsegraining length $l$. We present separately the high-temperature phase when the bath temperature is above the critical temperature ($T > T_\rmc$) from the low-temperature phase ($T < T_\rmc$). For each value of $l$, the corresponding equilibrium potential $\mathcal{V}_{\rm EQ}(\varphi)$ is also plotted for comparison (dotted lines).

In the high-temperature phase, see Fig.~\ref{fig:MC-NESS}~(a), the effect of reheating on the potential is found to be minimal. The equilibrium parabola is replaced by a potential with very similar features: a single global minimum at $\varphi=0$ and no other local minimum. This is all the more true when increasing the coarsegraining length $l$, \textit{i.e.} deeper in the infrared. This numerically validates the claim made around Eq.~(\ref{eq:VNESS_HT}) that above $T_\rmc$, the infrared fixed point is the equilibrium fixed point.

In stark contrast, the effect of reheating is found to be much more disruptive in the low-temperature phase ($T < T_\rmc$), see Fig.~\ref{fig:MC-NESS}~(b). There, the equilibrium Mexican-hat potential with two degenerate global minima is replaced by a potential $\mathcal{V}_{\rm NESS}(\varphi)$ with up to three minima: two degenerate minima at $\varphi \neq 0$, remnant features of the Mexican hat, and a new minimum at $\varphi=0$ which becomes the global minimum in the infrared.
Moreover, contrary to $\mathcal{V}_{\rm EQ}(\varphi)$, we verify that $\mathcal{V}_{\rm NESS}(\varphi)$ cannot be satisfactorily fitted with an (analytic) even polynomial of the form $\mbox{const} + a_2 \varphi^2 + a_4 \varphi^4 + a_6 \varphi^6 + a_8 \varphi^8$. Instead, numerical fits to the non-analytic form $\mathcal{V}_{\rm NESS}(\varphi) = \mbox{const} + c_1 |\varphi| + a_2 \varphi^2 + c_3 |\varphi|^3 + a_4 \varphi^4$ give excellent results. Details of the fitting procedure are given in Appendix~\ref{app:fit}. This numerically validates the claims made around Eq.~(\ref{eq:VNESS_LT}) that below $T_\rmc$ the Landau potential is non-analytic and flows towards a non-trivial non-equilibrium fixed point.
As an example, let us give the result of such a fitting procedure on the potential $\mathcal{V}_{\rm NESS}(\varphi)$ plotted in Fig.~\ref{fig:MC-NESS}~(b): We find $c_1(l=8)/c_1(l=4) \approx 1.9 > 1$, illustrating that the non-analytic operator $|\varphi|$ is infrared relevant below $T_\rmc$.

\paragraph{Validating the mean-field approximation} \label{sec:FP}

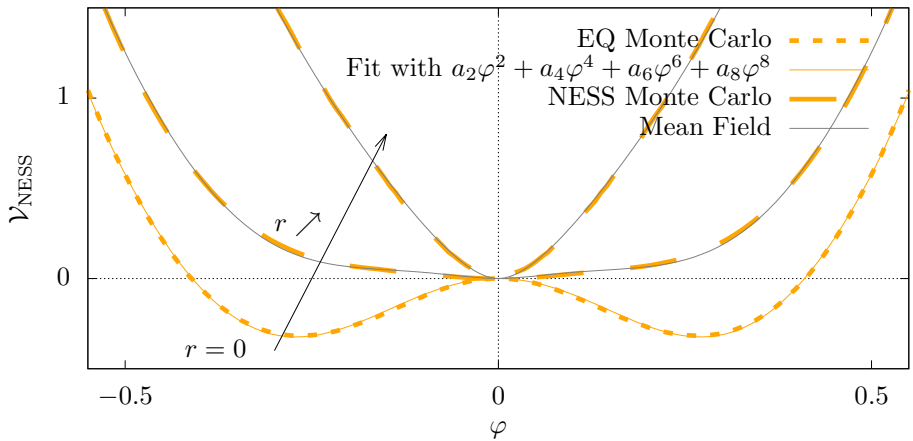
\begin{figure}
\center
\input{FP.tex}
\caption{\label{fig:FP} Validity of the mean-field approximation in $d=4$. Shown is a comparison non-equilibrium steady-state potentials $\mathcal{V}_{\rm NESS}(\varphi)$ computed from single-particle Langevin dynamics in Eq.~(\ref{eq:Langevin}) and $\eta = 1920$ (solid black lines) with those obtained from full-fledged Monte Carlo numerics (dashed lines) and for various reheating rates: $r=0,\, 0.005,\, 0.05$. The $r=0$ data corresponds to the equilibrium potential $\mathcal{V}_{\rm EQ}(\varphi)$. It was fitted with an (analytic) even potential and used as an input to the mean-field computation. The other parameters are 
$L = 8$, $l = 4$, and $T = 0.99\, T_\rmc$.
}
\end{figure}

We also numerically check the mean-field approximation that was used when simplifying the Model A dynamics in Eq.~(\ref{eq:modelA}) with the single-particle Langevin dynamics in Eq.~(\ref{eq:Langevin}). 
For that purpose, we compare the results of the full-fledged Monte Carlo numerics presented above with those obtained from solving the Langevin dynamics.
In practice, the equilibrium potential $\mathcal{V}_{\rm EQ}(\varphi)$ is numerically extracted from \emph{equilibrium} Monte Carlo dynamics and fitted with an (analytic) even polynomial. We illustrate the quality of such fits in Fig.~\ref{fig:FP}.
Then, the Fokker-Planck equation corresponding to Eq.~(\ref{eq:Langevin}), namely
\begin{align}
\eta \partial_t P_0(\varphi ; t) = \partial_\varphi \left[ \partial_\varphi \mathcal{V}_{\rm EQ}(\varphi) \, P_0(\varphi ; t) + \partial_\varphi P_0(\varphi ; t) \right] \,,
\end{align}
with the initial condition $P( \varphi ; t=0) = \delta(\varphi)$, is solved numerically. 
The ``friction'' parameter $\eta$ can be determined by requiring that the observable $\langle |\varphi(t)| \rangle$ computed within this framework matches the one computed from full-fledged Monte Carlo numerics (see details in Appendix~\ref{app:eta}).
Finally, the non-equilibrium steady-state potential in the presence of a finite reheating rate, $\mathcal{V}_{\rm NESS}$, is computed with the use of the renewal formula in Eq.~(\ref{eq:renewal}).

In Fig.~\ref{fig:FP}, we compare the resulting mean-field potentials to the ones obtained with full-fledged Monte Carlo numerics. The agreement is very good, thus validating the use of the mean-field approximation above the upper critical dimension and, therefore, the reduction of the non-equilibrium many-body problem to the non-equilibrium dynamics of a single particle in an \textit{ad-hoc} environment.

If it were necessary, this also constitutes an \textit{a posteriori} check that the Model A dynamics in Eq.~(\ref{eq:modelA}), involving the equilibrium free energy $\mathcal{F}_{\rm EQ}[\varphi]$ given in Eq.~(\ref{eq:F_eq}), are a faithful representation of the relaxation dynamics of the Ising model after a quench from infinite temperature\footnote{The match between Model A and Monte Carlo dynamics is only expected to hold at lengthscales larger than the microscopic lattice spacing and timescales larger than a Monte Carlo step. 
}.

\section{Below the upper critical dimension and open questions} \label{sec:discussion}

In dimensions below the upper critical dimension, $d_{\rm uc} = 4$, the spatial fluctuations of the field cannot be neglected, signaling the breakdown of the validity of the mean-field approximation. At equilibrium, the $O(n=1)$-symmetric phase is still Gaussian but the broken-symmetry phase is now characterized by the so-called Wilson-Fisher fixed point where $\lambda$ is finite. For finite reheating rates, the general comments made in Sec.~\ref{sec:simple} still apply: We expect a destabilization of the Wilson-Fisher fixed point but the involvement of non-analytic operators down to the infrared limit is yet to be confirmed.

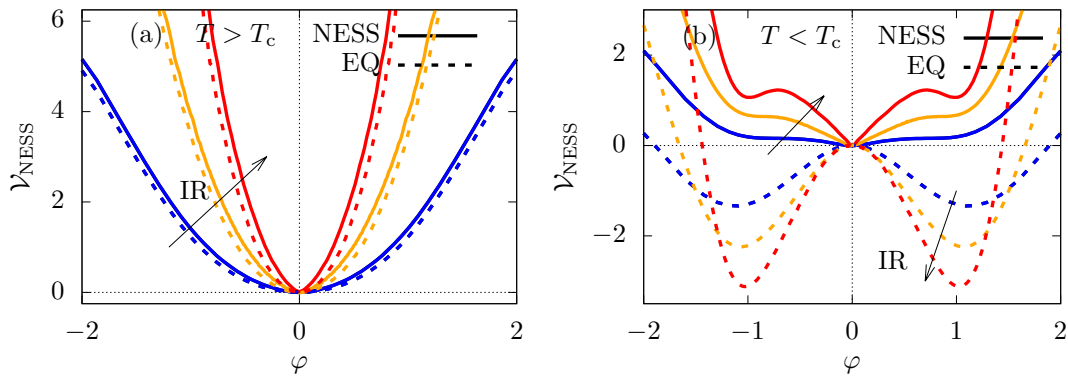
\begin{figure}
\center
\input{RG_HT_d2.tex}
\hspace{-2em}
\input{RG_LT_d2.tex}
\caption{\label{fig:MC-NESS_d2}
The $d=2$ Monte Carlo solution of the non-equilibrium steady-state potential $\mathcal{V}_{\rm NESS}(\varphi)$ for increasing coarsegraining lengths: $l = 8,\, 16,\, 32$ ($L=32$, $r=0.005$).
(a) Above $T_\rmc$ ($T = 1.1 \, T_\rmc$), $\mathcal{V}_{\rm NESS}$ (solid lines) does not differ much from the equilibrium potential $\mathcal{V}_{\rm EQ}$ (dotted lines).
(b) Below $T_\rmc$ ($T = 0.99 \, T_\rmc$), $\mathcal{V}_{\rm NESS}$ deviates markedly from the equilibrium Mexican hat. A cusp develops at the new minimum $\varphi=0$ which becomes the global minimum in the infrared (IR).}
\end{figure}

\paragraph{Monte Carlo numerics in $d=2$}
Given the lack of proper analytic methodology to perform such an RG computation, we may seek answers from Monte Carlo numerics in $d=2$. The details of the numerics are the same as for the $d=4$ case presented above. 
We also stay slightly away from the critical temperature $T_\rmc(d=2) = 2/\log(1+\sqrt{2}) \approx 2.27$ by working with $T = 0.99 \, T_\rmc$ and $T = 1.1 \, T_\rmc$.
Altogether, the results are qualitatively similar to those that were obtained in $d=4$.
In Fig.~\ref{fig:MC-NESS_d2}, the non-equilibrium steady-state potential $V_{\rm NESS}(\varphi)$ is plotted for different values of the coarsegraining length $l$.

Above $T_\rmc$, see Fig.~\ref{fig:MC-NESS_d2}~(a), the potential is similar to the equilibrium potential, with a single minimum at $\varphi = 0$, and an overall quadratic shape. Contrary to the $d=4$ case in Fig.~\ref{fig:MC-NESS}~(a), we may still note that the potential is slightly steeper in the presence of stochastic reheating. Additional investigations are needed to determine whether this difference subsists deeper in the infrared.

Below $T_\rmc$, see Fig.~\ref{fig:MC-NESS_d2}~(b), the potential develops a new minimum at $\varphi=0$ with the same characteristic cusp as the one previously attributed to the non-analytic operator $|\varphi|$ in $d=4$. As predicted in Sec.~\ref{sec:simple}, and similarly to the $d=4$ case, this minimum becomes the global minimum in the infrared. This strongly support the scenario of the equilibrium Wilson-Fisher fixed point being destabilized by non-analytic operators that are infrared relevant.

\paragraph{Non-analytic gradient terms}

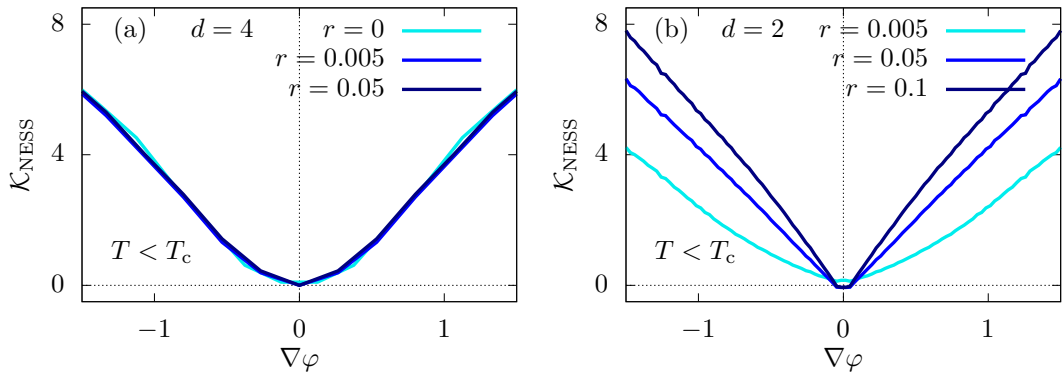
\begin{figure}
\center
\input{Grad_LT_d4.tex}
\hspace{-2em}
\input{Grad_LT_d2.tex}
\caption{\label{fig:gradients}
Monte Carlo solution of the non-equilibrium steady-state ``gradient potential'' $\mathcal{K}_{\rm NESS}(\nabla \varphi) \equiv - \log P_{\rm NESS}(\nabla \varphi)$ for various reheating rates $r$ given in the legend. (a)~In $d = 4$: $\mathcal{K}_{\rm NESS}$ does not appear to differ from equilibrium ($r=0$) [$L = 8$, $l = 2$, $T = 0.99\,T_\rmc$].
(b)~In $d=2$: $\mathcal{K}_{\rm NESS}$ undoubtedly depends on $r$ and could feature non-analyticities [$L = 64$, $l = 16$, $T = 0.99\,T_\rmc$].
}
\end{figure}

In addition to the non-analytic terms in the Landau potential, it is fair to wonder whether the non-equilibrium steady-state field theory can also feature non-analytic gradient terms of the type, \textit{e.g.}, $|\nabla \varphi|$, or $|\varphi| (\nabla \varphi)^2$. They would also be of intrinsic non-equilibrium nature and would depend on the reheating rate. Under RG, they could be generated by those non-analyticities already present in the potential. 

Above the upper critical dimension, in the regime of validity of the mean-field approximation, gradient terms are expected to be inconsequential and the reheating protocol is not expected to alter this picture. To check this, we numerically compute the probability distribution of gradients on the non-equilibrium steady state, $P_{\rm NESS}(\nabla \varphi)$. We choose a relatively small coarsegraining length $l$ to operate before the infrared fixed point is reached. In Fig.~\ref{fig:gradients}~(a), we plot the ``gradient potential'' $\mathcal{K}_{\rm NESS}(\nabla \varphi) \equiv - \log P_{\rm NESS}(\nabla \varphi)$ for various values of the reheating rate $r$ (including the equilibrium case $r=0$), both below and above $T_\rmc$. $\mathcal{K}_{\rm NESS}(\nabla \varphi)$ clearly appears to be independent of $r$, supporting the scenario that non-equilibrium reheating dynamics do not affect the gradient terms of the equilibrium Landau-Ginzburg-Wilson theory. This justifies the form of the effective Landau-Ginzburg functional that was proposed in Eq.~(\ref{eq:F_NESS}), with an analytic gradient expansion starting with a $(\nabla \varphi)^2$ term.

Preliminary results suggest that the scenario may be different below the upper critical dimension. In $d=2$, $\mathcal{K}_{\rm NESS}(\nabla \varphi)$ is found to depend on the reheating rate $r$, and is quite different from the equilibrium case. We illustrate this point in Fig.~\ref{fig:gradients}~(b). Whether this will survive in the infrared, and whether this does involve non-analyticities in the gradient terms are left for future investigations.

\paragraph{RG approach}
Ultimately, the discussion of the non-analytic non-equilibrium field theory below the upper critical dimension requires the construction of a computational framework to perform RG calculations. One route, specific to the reheating dynamics, is to use the renewal formula~(\ref{eq:renewal}). A time-dependent RG computation could, in principle, be performed at the level of the relaxation dynamics after a quench and the results translated to the case of reheating dynamics.
A more ambitious route would be to start directly from a non-analytic Landau-Ginzburg functional such as the one in Eq.~(\ref{eq:F_NESS}) and identify practical ways to perform the perturbative integration of high-energy modes in the presence of terms such as $|\varphi|$. 

\section*{Acknowledgements}
M.~K. is grateful to the \'Ecole Normale Sup\'erieure in Paris, where this work was initiated.
We are grateful to Claudio Chamon, Leticia Cugliandolo, Chandan Dasgupta, Satya Majumdar and Gregory Schehr for insightful discussions and for pointing out relevant literature.

\paragraph{Funding information}
This work benefited from the support of the ANR project MOMA (C.A.) and of the CEFIPRA Project No. 6004-1 (C.A. and M.K.). M.K. acknowledges Ramanujan Fellowship (Grant No.~SB/S2/RJN-114/2016), Early Career Research Award (Grant No.~ECR/2018/002085) and Matrics Grant (No.~MTR/2019/001101) from the Science and Engineering Research Board (SERB), Department of Science and Technology, Government of India.

\begin{appendix}

\section{Fitting the potential by a non-analytic function}
\label{app:fit}
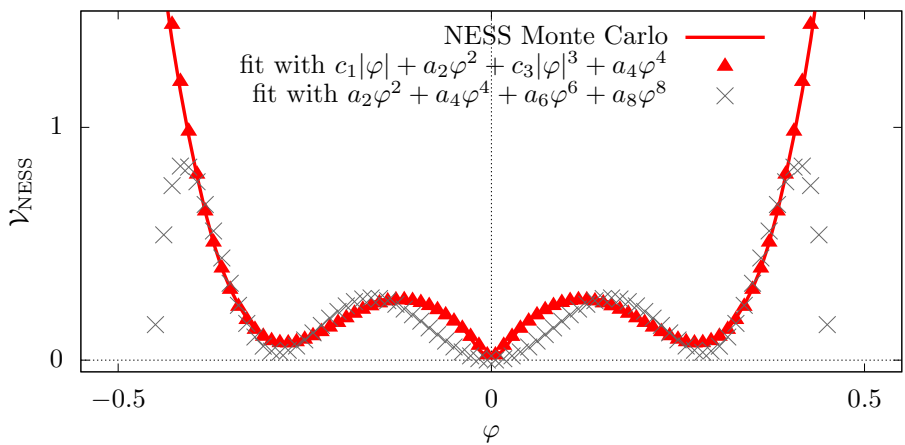
\begin{figure}[h]
\center
\input{fits.tex}
\caption{\label{fig:fits} The potential $\mathcal{V}_{\rm NESS}(\varphi)$ obtained from Monte Carlo numerics in $d=4$ for $T= 0.99 \, T_\rmc$, $r=0.005$ and $L=l=32$ [see the solid red curve in Fig.~\ref{fig:MC-NESS}~(b)] is fitted with both a non-analytic form (triangles) and an analytic form (crosses).}
\end{figure}
All the non-equilibrium steady states potentials $\mathcal{V}_{\rm NESS}$, at finite reheating rate $r > 0$, computed by means of Monte Carlo numerics and presented in this paper can be successfully fitted with a non analytic function of the form
$ \mbox{const} + c_1 |\varphi| + a_2 \varphi^2 + c_3 |\varphi|^3 + a_4 \varphi^4 + \ldots\,.$
In Fig.~\ref{fig:fits}, we provide an example of a fit of the potential $\mathcal{V}_{\rm NESS}(\varphi)$ presented in Fig.~\ref{fig:MC-NESS}~(b) at $T= 0.99 \, T_\rmc$, $r=0.005$ and $L=l=32$ and reproduced in Fig.~\ref{fig:fits} with a solid line. After setting the overall additive constant such that $\mathcal{V}_{\rm NESS}(0)=0$, it is fitted with a non-analytic function of the form
\begin{align} \label{eq:non-anal-function}
c_1 |\varphi| + a_2 \varphi^2 + c_3 |\varphi|^3 + a_4 \varphi^4\,,
\end{align}
where $c_1$, $a_2$, $c_3$ and $a_4$ are four fitting parameters determined by a nonlinear least-squares Levenberg-Marquardt algorithm. The resulting fit, represented by triangles, is excellent.
For comparison, the crosses show the result of a fit to the analytic function
\begin{align} \label{eq:anal-function}
 a_2 \varphi^2 + a_4 \varphi^4 + a_6 \varphi^6 + a_8 \varphi^8\,,
\end{align}
where $a_2$, $a_4$, $a_6$, and $a_8$ are also four fitting parameters. This fit fails to reproduce the data especially close to the cusp at $\varphi = 0$.

\section{Determining the coefficient $\eta$ of Model A dynamics} \label{app:eta}

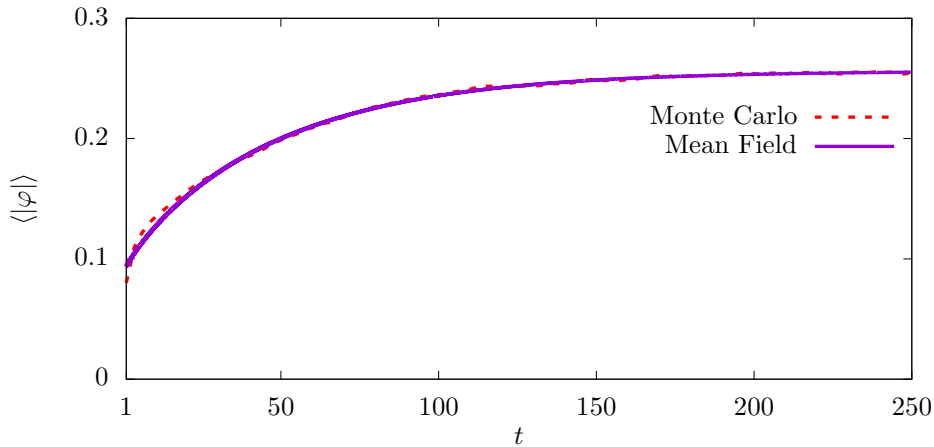
\begin{figure}[h]
\center
\input{Mag_t.tex}
\caption{\label{fig:etafit} Agreement of the time-dependent observable $\langle | \varphi| \rangle$ computed both from full-fledged Monte Carlo numerics in $d=4$ after a quench from infinite temperature ($L=8,\, l=4,\, T = 0.99 \, T_\rmc $), and from the mean-field Langevin approach with the initial condition $\varphi(t=0)=0$ and the equilibrium potential $\mathcal{V}_{\rm EQ}(\varphi)$ fitted in Fig.~\ref{fig:FP}. The dynamic coefficient $\eta$ entering the Langevin equation~(\ref{eq:Langevin}) has been set to $\eta \approx 1920$ for the two curves to match.}
\end{figure}
The dynamic ``friction'' parameter $\eta$ that enters the Model A dynamics in Eq.~(\ref{eq:modelA}), or its mean-field version in the Langevin dynamics of Eq.~(\ref{eq:Langevin}), is a scale dependent parameter. We determine its value by requiring the relaxation dynamics of a simple observable, namely (the absolute value of) the coarsegrained magnetization after a quench from infinite temperature, to be the same when computed from the full-fledged Monte Carlo dynamics in $d=4$ and when computed from the mean-field Langevin approach.
In practice, this means finding the value of $\eta$ such that 
\begin{align}
 \langle | \phi^{\rm MF} (t)| \rangle_{\xi} &= \int_{-\infty}^\infty \ud{\varphi} |\tanh \varphi | \, P_0(\varphi ; t) \,,
\end{align}
where $\langle \ldots \rangle_\xi$ indicates the average with respect to the Langevin noise, matches 
\begin{align}
\langle \big{|} \phi^{\rm MC} (x,t) \big{|} \rangle_{x, {\rm MC}} = \langle \Big{|} \frac{1}{l^d} \sum_{i \in v_l(x)} S_i(t) \Big{|} \rangle_{x, {\rm MC}} \,,
\end{align}
where $ \langle \ldots \rangle_{x, {\rm MC}}$ indicates the average with respect to the position and the Monte Carlo realizations and we recall the field redefinition $\varphi \equiv \mbox{arctanh}(\phi)$ .

We give an example of such a fit in Fig.~\ref{fig:etafit}. The excellent agreement between the two methods is another indication that the Model A dynamics in Eq.~(\ref{eq:modelA}), involving the equilibrium free energy $\mathcal{F}_{\rm EQ}[\varphi]$ given in Eq.~(\ref{eq:F_eq}), are a faithful representation of the relaxation dynamics of the Ising model after a quench from infinite temperature.

\end{appendix}

\nolinenumbers

\end{document}

%% file: FxPt.tex
\begingroup
  \makeatletter
  \providecommand\color[2][]{%
    \GenericError{(gnuplot) \space\space\space\@spaces}{%
      Package color not loaded in conjunction with
      terminal option `colourtext'%
    }{See the gnuplot documentation for explanation.%
    }{Either use 'blacktext' in gnuplot or load the package
      color.sty in LaTeX.}%
    \renewcommand\color[2][]{}%
  }%
  \providecommand\includegraphics[2][]{%
    \GenericError{(gnuplot) \space\space\space\@spaces}{%
      Package graphicx or graphics not loaded%
    }{See the gnuplot documentation for explanation.%
    }{The gnuplot epslatex terminal needs graphicx.sty or graphics.sty.}%
    \renewcommand\includegraphics[2][]{}%
  }%
  \providecommand\rotatebox[2]{#2}%
  \@ifundefined{ifGPcolor}{%
    \newif\ifGPcolor
    \GPcolorfalse
  }{}%
  \@ifundefined{ifGPblacktext}{%
    \newif\ifGPblacktext
    \GPblacktexttrue
  }{}%
  \let\gplgaddtomacro\g@addto@macro
  \gdef\gplbacktext{}%
  \gdef\gplfronttext{}%
  \makeatother
  \ifGPblacktext
    \def\colorrgb#1{}%
    \def\colorgray#1{}%
  \else
    \ifGPcolor
      \def\colorrgb#1{\color[rgb]{#1}}%
      \def\colorgray#1{\color[gray]{#1}}%
      \expandafter\def\csname LTw\endcsname{\color{white}}%
      \expandafter\def\csname LTb\endcsname{\color{black}}%
      \expandafter\def\csname LTa\endcsname{\color{black}}%
      \expandafter\def\csname LT0\endcsname{\color[rgb]{1,0,0}}%
      \expandafter\def\csname LT1\endcsname{\color[rgb]{0,1,0}}%
      \expandafter\def\csname LT2\endcsname{\color[rgb]{0,0,1}}%
      \expandafter\def\csname LT3\endcsname{\color[rgb]{1,0,1}}%
      \expandafter\def\csname LT4\endcsname{\color[rgb]{0,1,1}}%
      \expandafter\def\csname LT5\endcsname{\color[rgb]{1,1,0}}%
      \expandafter\def\csname LT6\endcsname{\color[rgb]{0,0,0}}%
      \expandafter\def\csname LT7\endcsname{\color[rgb]{1,0.3,0}}%
      \expandafter\def\csname LT8\endcsname{\color[rgb]{0.5,0.5,0.5}}%
    \else
      \def\colorrgb#1{\color{black}}%
      \def\colorgray#1{\color[gray]{#1}}%
      \expandafter\def\csname LTw\endcsname{\color{white}}%
      \expandafter\def\csname LTb\endcsname{\color{black}}%
      \expandafter\def\csname LTa\endcsname{\color{black}}%
      \expandafter\def\csname LT0\endcsname{\color{black}}%
      \expandafter\def\csname LT1\endcsname{\color{black}}%
      \expandafter\def\csname LT2\endcsname{\color{black}}%
      \expandafter\def\csname LT3\endcsname{\color{black}}%
      \expandafter\def\csname LT4\endcsname{\color{black}}%
      \expandafter\def\csname LT5\endcsname{\color{black}}%
      \expandafter\def\csname LT6\endcsname{\color{black}}%
      \expandafter\def\csname LT7\endcsname{\color{black}}%
      \expandafter\def\csname LT8\endcsname{\color{black}}%
    \fi
  \fi
    \setlength{\unitlength}{0.0500bp}%
    \ifx\gptboxheight\undefined%
      \newlength{\gptboxheight}%
      \newlength{\gptboxwidth}%
      \newsavebox{\gptboxtext}%
    \fi%
    \setlength{\fboxrule}{0.5pt}%
    \setlength{\fboxsep}{1pt}%
\begin{picture}(4320.00,3024.00)%
    \gplgaddtomacro\gplbacktext{%
      \csname LTb\endcsname
      \put(550,910){\makebox(0,0)[r]{\strut{}$0$}}%
      \put(550,2172){\makebox(0,0)[r]{\strut{}$1$}}%
      \put(759,374){\makebox(0,0){\strut{}$-2$}}%
      \put(1531,374){\makebox(0,0){\strut{}$-1$}}%
      \put(2303,374){\makebox(0,0){\strut{}$0$}}%
      \put(3074,374){\makebox(0,0){\strut{}$1$}}%
      \put(3846,374){\makebox(0,0){\strut{}$2$}}%
      \csname LTb\endcsname
      \put(759,2551){\makebox(0,0)[l]{\strut{}(a)}}%
    }%
    \gplgaddtomacro\gplfronttext{%
      \csname LTb\endcsname
      \put(185,1698){\rotatebox{-270}{\makebox(0,0){\strut{}$\mathcal{V}^*_{\rm NESS}$}}}%
      \put(2302,154){\makebox(0,0){\strut{}$\varPhi \equiv \sqrt{|\mu|} \varphi $}}%
      \put(2302,2693){\makebox(0,0){\strut{}}}%
      \csname LTb\endcsname
      \put(2296,2441){\makebox(0,0)[r]{\strut{}$T<T_{\rm c}$}}%
      \csname LTb\endcsname
      \put(2296,2221){\makebox(0,0)[r]{\strut{}$T>T_{\rm c}$}}%
    }%
    \gplbacktext
    \put(0,0){\includegraphics{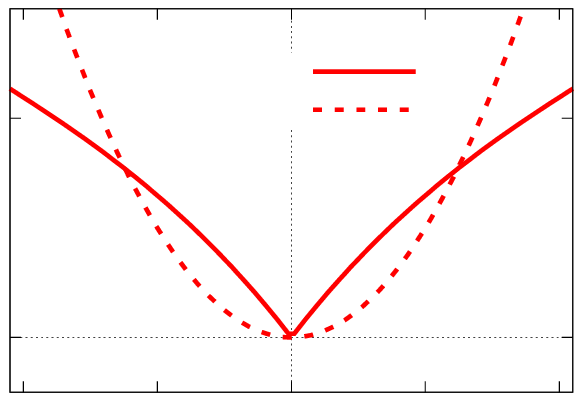}}%
    \gplfronttext
  \end{picture}%
\endgroup

%% file: FxPt_Tc.tex
\begingroup
  \makeatletter
  \providecommand\color[2][]{%
    \GenericError{(gnuplot) \space\space\space\@spaces}{%
      Package color not loaded in conjunction with
      terminal option `colourtext'%
    }{See the gnuplot documentation for explanation.%
    }{Either use 'blacktext' in gnuplot or load the package
      color.sty in LaTeX.}%
    \renewcommand\color[2][]{}%
  }%
  \providecommand\includegraphics[2][]{%
    \GenericError{(gnuplot) \space\space\space\@spaces}{%
      Package graphicx or graphics not loaded%
    }{See the gnuplot documentation for explanation.%
    }{The gnuplot epslatex terminal needs graphicx.sty or graphics.sty.}%
    \renewcommand\includegraphics[2][]{}%
  }%
  \providecommand\rotatebox[2]{#2}%
  \@ifundefined{ifGPcolor}{%
    \newif\ifGPcolor
    \GPcolorfalse
  }{}%
  \@ifundefined{ifGPblacktext}{%
    \newif\ifGPblacktext
    \GPblacktexttrue
  }{}%
  \let\gplgaddtomacro\g@addto@macro
  \gdef\gplbacktext{}%
  \gdef\gplfronttext{}%
  \makeatother
  \ifGPblacktext
    \def\colorrgb#1{}%
    \def\colorgray#1{}%
  \else
    \ifGPcolor
      \def\colorrgb#1{\color[rgb]{#1}}%
      \def\colorgray#1{\color[gray]{#1}}%
      \expandafter\def\csname LTw\endcsname{\color{white}}%
      \expandafter\def\csname LTb\endcsname{\color{black}}%
      \expandafter\def\csname LTa\endcsname{\color{black}}%
      \expandafter\def\csname LT0\endcsname{\color[rgb]{1,0,0}}%
      \expandafter\def\csname LT1\endcsname{\color[rgb]{0,1,0}}%
      \expandafter\def\csname LT2\endcsname{\color[rgb]{0,0,1}}%
      \expandafter\def\csname LT3\endcsname{\color[rgb]{1,0,1}}%
      \expandafter\def\csname LT4\endcsname{\color[rgb]{0,1,1}}%
      \expandafter\def\csname LT5\endcsname{\color[rgb]{1,1,0}}%
      \expandafter\def\csname LT6\endcsname{\color[rgb]{0,0,0}}%
      \expandafter\def\csname LT7\endcsname{\color[rgb]{1,0.3,0}}%
      \expandafter\def\csname LT8\endcsname{\color[rgb]{0.5,0.5,0.5}}%
    \else
      \def\colorrgb#1{\color{black}}%
      \def\colorgray#1{\color[gray]{#1}}%
      \expandafter\def\csname LTw\endcsname{\color{white}}%
      \expandafter\def\csname LTb\endcsname{\color{black}}%
      \expandafter\def\csname LTa\endcsname{\color{black}}%
      \expandafter\def\csname LT0\endcsname{\color{black}}%
      \expandafter\def\csname LT1\endcsname{\color{black}}%
      \expandafter\def\csname LT2\endcsname{\color{black}}%
      \expandafter\def\csname LT3\endcsname{\color{black}}%
      \expandafter\def\csname LT4\endcsname{\color{black}}%
      \expandafter\def\csname LT5\endcsname{\color{black}}%
      \expandafter\def\csname LT6\endcsname{\color{black}}%
      \expandafter\def\csname LT7\endcsname{\color{black}}%
      \expandafter\def\csname LT8\endcsname{\color{black}}%
    \fi
  \fi
    \setlength{\unitlength}{0.0500bp}%
    \ifx\gptboxheight\undefined%
      \newlength{\gptboxheight}%
      \newlength{\gptboxwidth}%
      \newsavebox{\gptboxtext}%
    \fi%
    \setlength{\fboxrule}{0.5pt}%
    \setlength{\fboxsep}{1pt}%
\begin{picture}(4320.00,3024.00)%
    \gplgaddtomacro\gplbacktext{%
      \csname LTb\endcsname
      \put(550,910){\makebox(0,0)[r]{\strut{}$0$}}%
      \put(550,2172){\makebox(0,0)[r]{\strut{}$1$}}%
      \put(759,374){\makebox(0,0){\strut{}$-2$}}%
      \put(1531,374){\makebox(0,0){\strut{}$-1$}}%
      \put(2303,374){\makebox(0,0){\strut{}$0$}}%
      \put(3074,374){\makebox(0,0){\strut{}$1$}}%
      \put(3846,374){\makebox(0,0){\strut{}$2$}}%
      \csname LTb\endcsname
      \put(759,2551){\makebox(0,0)[l]{\strut{}(b)}}%
    }%
    \gplgaddtomacro\gplfronttext{%
      \csname LTb\endcsname
      \put(185,1698){\rotatebox{-270}{\makebox(0,0){\strut{}$\mathcal{V}^*_{\rm NESS}$}}}%
      \put(2302,154){\makebox(0,0){\strut{}$\sqrt{r} \times ( \varPhi \equiv \sqrt{\eta} \varphi) $}}%
      \put(2302,2693){\makebox(0,0){\strut{}}}%
      \csname LTb\endcsname
      \put(2296,2441){\makebox(0,0)[r]{\strut{}$T=T_{\rm c}$}}%
    }%
    \gplbacktext
    \put(0,0){\includegraphics{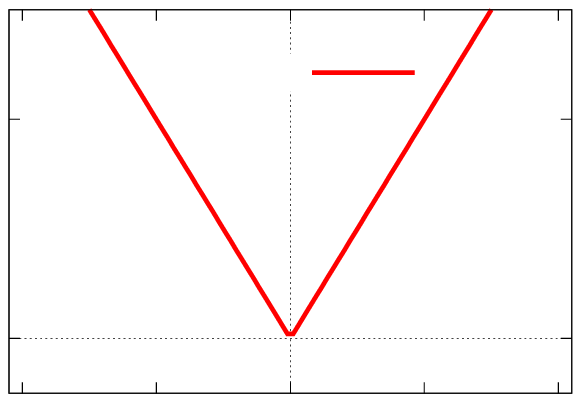}}%
    \gplfronttext
  \end{picture}%
\endgroup

%% file: RG_HT.tex
\begingroup
  \makeatletter
  \providecommand\color[2][]{%
    \GenericError{(gnuplot) \space\space\space\@spaces}{%
      Package color not loaded in conjunction with
      terminal option `colourtext'%
    }{See the gnuplot documentation for explanation.%
    }{Either use 'blacktext' in gnuplot or load the package
      color.sty in LaTeX.}%
    \renewcommand\color[2][]{}%
  }%
  \providecommand\includegraphics[2][]{%
    \GenericError{(gnuplot) \space\space\space\@spaces}{%
      Package graphicx or graphics not loaded%
    }{See the gnuplot documentation for explanation.%
    }{The gnuplot epslatex terminal needs graphicx.sty or graphics.sty.}%
    \renewcommand\includegraphics[2][]{}%
  }%
  \providecommand\rotatebox[2]{#2}%
  \@ifundefined{ifGPcolor}{%
    \newif\ifGPcolor
    \GPcolorfalse
  }{}%
  \@ifundefined{ifGPblacktext}{%
    \newif\ifGPblacktext
    \GPblacktexttrue
  }{}%
  \let\gplgaddtomacro\g@addto@macro
  \gdef\gplbacktext{}%
  \gdef\gplfronttext{}%
  \makeatother
  \ifGPblacktext
    \def\colorrgb#1{}%
    \def\colorgray#1{}%
  \else
    \ifGPcolor
      \def\colorrgb#1{\color[rgb]{#1}}%
      \def\colorgray#1{\color[gray]{#1}}%
      \expandafter\def\csname LTw\endcsname{\color{white}}%
      \expandafter\def\csname LTb\endcsname{\color{black}}%
      \expandafter\def\csname LTa\endcsname{\color{black}}%
      \expandafter\def\csname LT0\endcsname{\color[rgb]{1,0,0}}%
      \expandafter\def\csname LT1\endcsname{\color[rgb]{0,1,0}}%
      \expandafter\def\csname LT2\endcsname{\color[rgb]{0,0,1}}%
      \expandafter\def\csname LT3\endcsname{\color[rgb]{1,0,1}}%
      \expandafter\def\csname LT4\endcsname{\color[rgb]{0,1,1}}%
      \expandafter\def\csname LT5\endcsname{\color[rgb]{1,1,0}}%
      \expandafter\def\csname LT6\endcsname{\color[rgb]{0,0,0}}%
      \expandafter\def\csname LT7\endcsname{\color[rgb]{1,0.3,0}}%
      \expandafter\def\csname LT8\endcsname{\color[rgb]{0.5,0.5,0.5}}%
    \else
      \def\colorrgb#1{\color{black}}%
      \def\colorgray#1{\color[gray]{#1}}%
      \expandafter\def\csname LTw\endcsname{\color{white}}%
      \expandafter\def\csname LTb\endcsname{\color{black}}%
      \expandafter\def\csname LTa\endcsname{\color{black}}%
      \expandafter\def\csname LT0\endcsname{\color{black}}%
      \expandafter\def\csname LT1\endcsname{\color{black}}%
      \expandafter\def\csname LT2\endcsname{\color{black}}%
      \expandafter\def\csname LT3\endcsname{\color{black}}%
      \expandafter\def\csname LT4\endcsname{\color{black}}%
      \expandafter\def\csname LT5\endcsname{\color{black}}%
      \expandafter\def\csname LT6\endcsname{\color{black}}%
      \expandafter\def\csname LT7\endcsname{\color{black}}%
      \expandafter\def\csname LT8\endcsname{\color{black}}%
    \fi
  \fi
    \setlength{\unitlength}{0.0500bp}%
    \ifx\gptboxheight\undefined%
      \newlength{\gptboxheight}%
      \newlength{\gptboxwidth}%
      \newsavebox{\gptboxtext}%
    \fi%
    \setlength{\fboxrule}{0.5pt}%
    \setlength{\fboxsep}{1pt}%
\begin{picture}(4320.00,3024.00)%
    \gplgaddtomacro\gplbacktext{%
      \csname LTb\endcsname
      \put(550,799){\makebox(0,0)[r]{\strut{}$0$}}%
      \put(550,1617){\makebox(0,0)[r]{\strut{}$1$}}%
      \put(550,2435){\makebox(0,0)[r]{\strut{}$2$}}%
      \put(829,374){\makebox(0,0){\strut{}$-0.5$}}%
      \put(2303,374){\makebox(0,0){\strut{}$0$}}%
      \put(3776,374){\makebox(0,0){\strut{}$0.5$}}%
      \csname LTb\endcsname
      \put(1448,1429){\makebox(0,0)[l]{\strut{}IR}}%
    }%
    \gplgaddtomacro\gplfronttext{%
      \csname LTb\endcsname
      \put(237,1698){\rotatebox{-270}{\makebox(0,0){\strut{}$\mathcal{V}_{\rm NESS}$}}}%
      \put(2302,154){\makebox(0,0){\strut{}$\varphi$}}%
      \put(2302,2693){\makebox(0,0){\strut{}}}%
      \csname LTb\endcsname
      \put(2921,2570){\makebox(0,0)[r]{\strut{}(a)   $\hspace{1ex} T > T_{\rm c}\hspace{6ex} $  NESS}}%
      \csname LTb\endcsname
      \put(2921,2350){\makebox(0,0)[r]{\strut{}EQ}}%
    }%
    \gplbacktext
    \put(0,0){\includegraphics{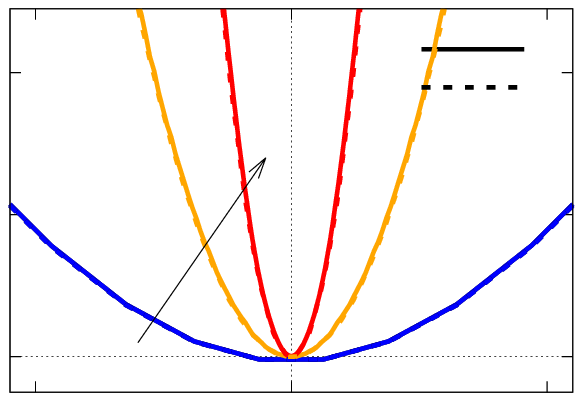}}%
    \gplfronttext
  \end{picture}%
\endgroup

%% file: RG_LT.tex
\begingroup
  \makeatletter
  \providecommand\color[2][]{%
    \GenericError{(gnuplot) \space\space\space\@spaces}{%
      Package color not loaded in conjunction with
      terminal option `colourtext'%
    }{See the gnuplot documentation for explanation.%
    }{Either use 'blacktext' in gnuplot or load the package
      color.sty in LaTeX.}%
    \renewcommand\color[2][]{}%
  }%
  \providecommand\includegraphics[2][]{%
    \GenericError{(gnuplot) \space\space\space\@spaces}{%
      Package graphicx or graphics not loaded%
    }{See the gnuplot documentation for explanation.%
    }{The gnuplot epslatex terminal needs graphicx.sty or graphics.sty.}%
    \renewcommand\includegraphics[2][]{}%
  }%
  \providecommand\rotatebox[2]{#2}%
  \@ifundefined{ifGPcolor}{%
    \newif\ifGPcolor
    \GPcolorfalse
  }{}%
  \@ifundefined{ifGPblacktext}{%
    \newif\ifGPblacktext
    \GPblacktexttrue
  }{}%
  \let\gplgaddtomacro\g@addto@macro
  \gdef\gplbacktext{}%
  \gdef\gplfronttext{}%
  \makeatother
  \ifGPblacktext
    \def\colorrgb#1{}%
    \def\colorgray#1{}%
  \else
    \ifGPcolor
      \def\colorrgb#1{\color[rgb]{#1}}%
      \def\colorgray#1{\color[gray]{#1}}%
      \expandafter\def\csname LTw\endcsname{\color{white}}%
      \expandafter\def\csname LTb\endcsname{\color{black}}%
      \expandafter\def\csname LTa\endcsname{\color{black}}%
      \expandafter\def\csname LT0\endcsname{\color[rgb]{1,0,0}}%
      \expandafter\def\csname LT1\endcsname{\color[rgb]{0,1,0}}%
      \expandafter\def\csname LT2\endcsname{\color[rgb]{0,0,1}}%
      \expandafter\def\csname LT3\endcsname{\color[rgb]{1,0,1}}%
      \expandafter\def\csname LT4\endcsname{\color[rgb]{0,1,1}}%
      \expandafter\def\csname LT5\endcsname{\color[rgb]{1,1,0}}%
      \expandafter\def\csname LT6\endcsname{\color[rgb]{0,0,0}}%
      \expandafter\def\csname LT7\endcsname{\color[rgb]{1,0.3,0}}%
      \expandafter\def\csname LT8\endcsname{\color[rgb]{0.5,0.5,0.5}}%
    \else
      \def\colorrgb#1{\color{black}}%
      \def\colorgray#1{\color[gray]{#1}}%
      \expandafter\def\csname LTw\endcsname{\color{white}}%
      \expandafter\def\csname LTb\endcsname{\color{black}}%
      \expandafter\def\csname LTa\endcsname{\color{black}}%
      \expandafter\def\csname LT0\endcsname{\color{black}}%
      \expandafter\def\csname LT1\endcsname{\color{black}}%
      \expandafter\def\csname LT2\endcsname{\color{black}}%
      \expandafter\def\csname LT3\endcsname{\color{black}}%
      \expandafter\def\csname LT4\endcsname{\color{black}}%
      \expandafter\def\csname LT5\endcsname{\color{black}}%
      \expandafter\def\csname LT6\endcsname{\color{black}}%
      \expandafter\def\csname LT7\endcsname{\color{black}}%
      \expandafter\def\csname LT8\endcsname{\color{black}}%
    \fi
  \fi
    \setlength{\unitlength}{0.0500bp}%
    \ifx\gptboxheight\undefined%
      \newlength{\gptboxheight}%
      \newlength{\gptboxwidth}%
      \newsavebox{\gptboxtext}%
    \fi%
    \setlength{\fboxrule}{0.5pt}%
    \setlength{\fboxsep}{1pt}%
\begin{picture}(4320.00,3024.00)%
    \gplgaddtomacro\gplbacktext{%
      \csname LTb\endcsname
      \put(682,758){\makebox(0,0)[r]{\strut{}$-1$}}%
      \put(682,1576){\makebox(0,0)[r]{\strut{}$0$}}%
      \put(682,2394){\makebox(0,0)[r]{\strut{}$1$}}%
      \put(955,374){\makebox(0,0){\strut{}$-0.5$}}%
      \put(2369,374){\makebox(0,0){\strut{}$0$}}%
      \put(3782,374){\makebox(0,0){\strut{}$0.5$}}%
      \csname LTb\endcsname
      \put(2877,1167){\makebox(0,0)[l]{\strut{}IR}}%
    }%
    \gplgaddtomacro\gplfronttext{%
      \csname LTb\endcsname
      \put(237,1698){\rotatebox{-270}{\makebox(0,0){\strut{}$\mathcal{V}_{\rm NESS}$}}}%
      \put(2368,154){\makebox(0,0){\strut{}$\varphi$}}%
      \put(2368,2693){\makebox(0,0){\strut{}}}%
      \csname LTb\endcsname
      \put(2927,2570){\makebox(0,0)[r]{\strut{}(b)  \hspace{1ex} $\ T < T_{\rm c}\ $  \hspace{1ex} NESS}}%
      \csname LTb\endcsname
      \put(2927,2350){\makebox(0,0)[r]{\strut{}EQ}}%
    }%
    \gplbacktext
    \put(0,0){\includegraphics{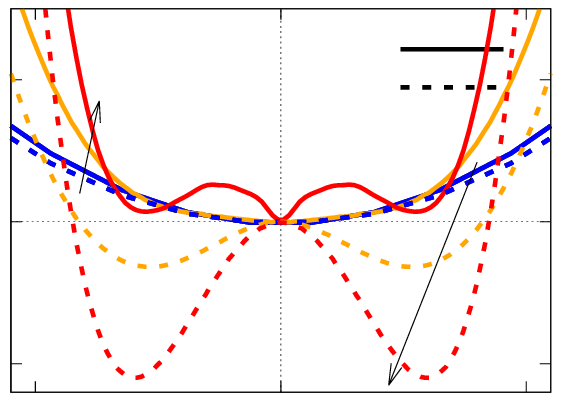}}%
    \gplfronttext
  \end{picture}%
\endgroup

%% file: FP.tex
\begingroup
  \makeatletter
  \providecommand\color[2][]{%
    \GenericError{(gnuplot) \space\space\space\@spaces}{%
      Package color not loaded in conjunction with
      terminal option `colourtext'%
    }{See the gnuplot documentation for explanation.%
    }{Either use 'blacktext' in gnuplot or load the package
      color.sty in LaTeX.}%
    \renewcommand\color[2][]{}%
  }%
  \providecommand\includegraphics[2][]{%
    \GenericError{(gnuplot) \space\space\space\@spaces}{%
      Package graphicx or graphics not loaded%
    }{See the gnuplot documentation for explanation.%
    }{The gnuplot epslatex terminal needs graphicx.sty or graphics.sty.}%
    \renewcommand\includegraphics[2][]{}%
  }%
  \providecommand\rotatebox[2]{#2}%
  \@ifundefined{ifGPcolor}{%
    \newif\ifGPcolor
    \GPcolorfalse
  }{}%
  \@ifundefined{ifGPblacktext}{%
    \newif\ifGPblacktext
    \GPblacktexttrue
  }{}%
  \let\gplgaddtomacro\g@addto@macro
  \gdef\gplbacktext{}%
  \gdef\gplfronttext{}%
  \makeatother
  \ifGPblacktext
    \def\colorrgb#1{}%
    \def\colorgray#1{}%
  \else
    \ifGPcolor
      \def\colorrgb#1{\color[rgb]{#1}}%
      \def\colorgray#1{\color[gray]{#1}}%
      \expandafter\def\csname LTw\endcsname{\color{white}}%
      \expandafter\def\csname LTb\endcsname{\color{black}}%
      \expandafter\def\csname LTa\endcsname{\color{black}}%
      \expandafter\def\csname LT0\endcsname{\color[rgb]{1,0,0}}%
      \expandafter\def\csname LT1\endcsname{\color[rgb]{0,1,0}}%
      \expandafter\def\csname LT2\endcsname{\color[rgb]{0,0,1}}%
      \expandafter\def\csname LT3\endcsname{\color[rgb]{1,0,1}}%
      \expandafter\def\csname LT4\endcsname{\color[rgb]{0,1,1}}%
      \expandafter\def\csname LT5\endcsname{\color[rgb]{1,1,0}}%
      \expandafter\def\csname LT6\endcsname{\color[rgb]{0,0,0}}%
      \expandafter\def\csname LT7\endcsname{\color[rgb]{1,0.3,0}}%
      \expandafter\def\csname LT8\endcsname{\color[rgb]{0.5,0.5,0.5}}%
    \else
      \def\colorrgb#1{\color{black}}%
      \def\colorgray#1{\color[gray]{#1}}%
      \expandafter\def\csname LTw\endcsname{\color{white}}%
      \expandafter\def\csname LTb\endcsname{\color{black}}%
      \expandafter\def\csname LTa\endcsname{\color{black}}%
      \expandafter\def\csname LT0\endcsname{\color{black}}%
      \expandafter\def\csname LT1\endcsname{\color{black}}%
      \expandafter\def\csname LT2\endcsname{\color{black}}%
      \expandafter\def\csname LT3\endcsname{\color{black}}%
      \expandafter\def\csname LT4\endcsname{\color{black}}%
      \expandafter\def\csname LT5\endcsname{\color{black}}%
      \expandafter\def\csname LT6\endcsname{\color{black}}%
      \expandafter\def\csname LT7\endcsname{\color{black}}%
      \expandafter\def\csname LT8\endcsname{\color{black}}%
    \fi
  \fi
    \setlength{\unitlength}{0.0500bp}%
    \ifx\gptboxheight\undefined%
      \newlength{\gptboxheight}%
      \newlength{\gptboxwidth}%
      \newsavebox{\gptboxtext}%
    \fi%
    \setlength{\fboxrule}{0.5pt}%
    \setlength{\fboxsep}{1pt}%
\begin{picture}(7200.00,3528.00)%
    \gplgaddtomacro\gplbacktext{%
      \csname LTb\endcsname
      \put(550,1272){\makebox(0,0)[r]{\strut{}$0$}}%
      \put(550,2629){\makebox(0,0)[r]{\strut{}$1$}}%
      \put(960,374){\makebox(0,0){\strut{}$-0.5$}}%
      \put(3743,374){\makebox(0,0){\strut{}$0$}}%
      \put(6525,374){\makebox(0,0){\strut{}$0.5$}}%
      \csname LTb\endcsname
      \put(2073,1679){\makebox(0,0)[l]{\strut{}$r \nearrow$}}%
      \put(1405,730){\makebox(0,0)[l]{\strut{}$r=0$}}%
    }%
    \gplgaddtomacro\gplfronttext{%
      \csname LTb\endcsname
      \put(185,1950){\rotatebox{-270}{\makebox(0,0){\strut{}$\mathcal{V}_{\rm NESS}$}}}%
      \put(3742,154){\makebox(0,0){\strut{}$\varphi$}}%
      \put(3742,3197){\makebox(0,0){\strut{}}}%
      \csname LTb\endcsname
      \put(5781,3061){\makebox(0,0)[r]{\strut{}EQ Monte Carlo}}%
      \csname LTb\endcsname
      \put(5781,2841){\makebox(0,0)[r]{\strut{}Fit with $a_2 \varphi^2 + a_4 \varphi^4 +  a_6 \varphi^6 + a_8 \varphi^8 $}}%
      \csname LTb\endcsname
      \put(5781,2621){\makebox(0,0)[r]{\strut{}NESS Monte Carlo}}%
      \csname LTb\endcsname
      \put(5781,2401){\makebox(0,0)[r]{\strut{}Mean Field}}%
    }%
    \gplbacktext
    \put(0,0){\includegraphics{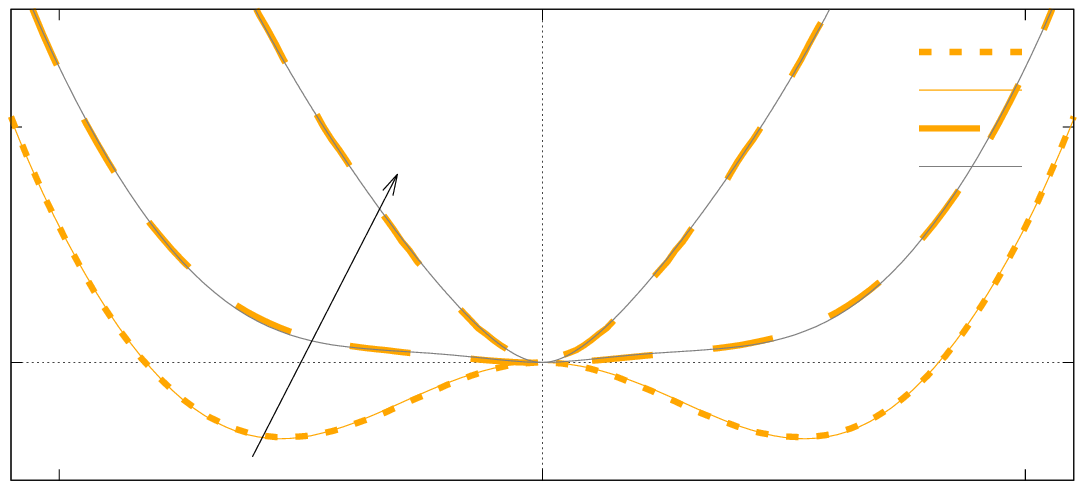}}%
    \gplfronttext
  \end{picture}%
\endgroup

%% file: RG_HT_d2.tex
\begingroup
  \makeatletter
  \providecommand\color[2][]{%
    \GenericError{(gnuplot) \space\space\space\@spaces}{%
      Package color not loaded in conjunction with
      terminal option `colourtext'%
    }{See the gnuplot documentation for explanation.%
    }{Either use 'blacktext' in gnuplot or load the package
      color.sty in LaTeX.}%
    \renewcommand\color[2][]{}%
  }%
  \providecommand\includegraphics[2][]{%
    \GenericError{(gnuplot) \space\space\space\@spaces}{%
      Package graphicx or graphics not loaded%
    }{See the gnuplot documentation for explanation.%
    }{The gnuplot epslatex terminal needs graphicx.sty or graphics.sty.}%
    \renewcommand\includegraphics[2][]{}%
  }%
  \providecommand\rotatebox[2]{#2}%
  \@ifundefined{ifGPcolor}{%
    \newif\ifGPcolor
    \GPcolorfalse
  }{}%
  \@ifundefined{ifGPblacktext}{%
    \newif\ifGPblacktext
    \GPblacktexttrue
  }{}%
  \let\gplgaddtomacro\g@addto@macro
  \gdef\gplbacktext{}%
  \gdef\gplfronttext{}%
  \makeatother
  \ifGPblacktext
    \def\colorrgb#1{}%
    \def\colorgray#1{}%
  \else
    \ifGPcolor
      \def\colorrgb#1{\color[rgb]{#1}}%
      \def\colorgray#1{\color[gray]{#1}}%
      \expandafter\def\csname LTw\endcsname{\color{white}}%
      \expandafter\def\csname LTb\endcsname{\color{black}}%
      \expandafter\def\csname LTa\endcsname{\color{black}}%
      \expandafter\def\csname LT0\endcsname{\color[rgb]{1,0,0}}%
      \expandafter\def\csname LT1\endcsname{\color[rgb]{0,1,0}}%
      \expandafter\def\csname LT2\endcsname{\color[rgb]{0,0,1}}%
      \expandafter\def\csname LT3\endcsname{\color[rgb]{1,0,1}}%
      \expandafter\def\csname LT4\endcsname{\color[rgb]{0,1,1}}%
      \expandafter\def\csname LT5\endcsname{\color[rgb]{1,1,0}}%
      \expandafter\def\csname LT6\endcsname{\color[rgb]{0,0,0}}%
      \expandafter\def\csname LT7\endcsname{\color[rgb]{1,0.3,0}}%
      \expandafter\def\csname LT8\endcsname{\color[rgb]{0.5,0.5,0.5}}%
    \else
      \def\colorrgb#1{\color{black}}%
      \def\colorgray#1{\color[gray]{#1}}%
      \expandafter\def\csname LTw\endcsname{\color{white}}%
      \expandafter\def\csname LTb\endcsname{\color{black}}%
      \expandafter\def\csname LTa\endcsname{\color{black}}%
      \expandafter\def\csname LT0\endcsname{\color{black}}%
      \expandafter\def\csname LT1\endcsname{\color{black}}%
      \expandafter\def\csname LT2\endcsname{\color{black}}%
      \expandafter\def\csname LT3\endcsname{\color{black}}%
      \expandafter\def\csname LT4\endcsname{\color{black}}%
      \expandafter\def\csname LT5\endcsname{\color{black}}%
      \expandafter\def\csname LT6\endcsname{\color{black}}%
      \expandafter\def\csname LT7\endcsname{\color{black}}%
      \expandafter\def\csname LT8\endcsname{\color{black}}%
    \fi
  \fi
    \setlength{\unitlength}{0.0500bp}%
    \ifx\gptboxheight\undefined%
      \newlength{\gptboxheight}%
      \newlength{\gptboxwidth}%
      \newsavebox{\gptboxtext}%
    \fi%
    \setlength{\fboxrule}{0.5pt}%
    \setlength{\fboxsep}{1pt}%
\begin{picture}(4320.00,3024.00)%
    \gplgaddtomacro\gplbacktext{%
      \csname LTb\endcsname
      \put(550,679){\makebox(0,0)[r]{\strut{}$0$}}%
      \put(550,1359){\makebox(0,0)[r]{\strut{}$2$}}%
      \put(550,2038){\makebox(0,0)[r]{\strut{}$4$}}%
      \put(550,2718){\makebox(0,0)[r]{\strut{}$6$}}%
      \put(682,374){\makebox(0,0){\strut{}$-2$}}%
      \put(2303,374){\makebox(0,0){\strut{}$0$}}%
      \put(3923,374){\makebox(0,0){\strut{}$2$}}%
      \csname LTb\endcsname
      \put(1411,1427){\makebox(0,0)[l]{\strut{}IR}}%
    }%
    \gplgaddtomacro\gplfronttext{%
      \csname LTb\endcsname
      \put(237,1698){\rotatebox{-270}{\makebox(0,0){\strut{}$\mathcal{V}_{\rm NESS}$}}}%
      \put(2302,154){\makebox(0,0){\strut{}$\varphi$}}%
      \put(2302,2693){\makebox(0,0){\strut{}}}%
      \csname LTb\endcsname
      \put(2906,2608){\makebox(0,0)[r]{\strut{}(a)   $\hspace{2ex} T > T_{\rm c}\hspace{1em} $  NESS}}%
      \csname LTb\endcsname
      \put(2906,2388){\makebox(0,0)[r]{\strut{}EQ}}%
    }%
    \gplbacktext
    \put(0,0){\includegraphics{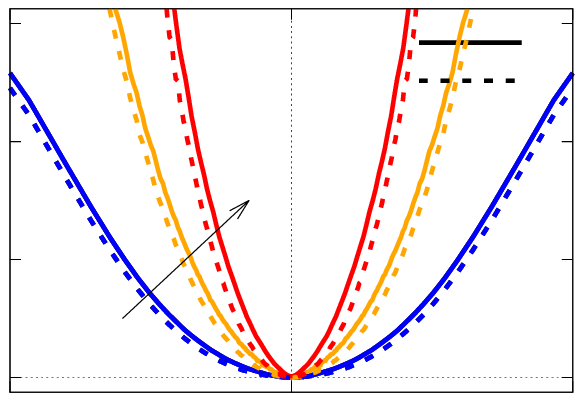}}%
    \gplfronttext
  \end{picture}%
\endgroup

%% file: RG_LT_d2.tex
\begingroup
  \makeatletter
  \providecommand\color[2][]{%
    \GenericError{(gnuplot) \space\space\space\@spaces}{%
      Package color not loaded in conjunction with
      terminal option `colourtext'%
    }{See the gnuplot documentation for explanation.%
    }{Either use 'blacktext' in gnuplot or load the package
      color.sty in LaTeX.}%
    \renewcommand\color[2][]{}%
  }%
  \providecommand\includegraphics[2][]{%
    \GenericError{(gnuplot) \space\space\space\@spaces}{%
      Package graphicx or graphics not loaded%
    }{See the gnuplot documentation for explanation.%
    }{The gnuplot epslatex terminal needs graphicx.sty or graphics.sty.}%
    \renewcommand\includegraphics[2][]{}%
  }%
  \providecommand\rotatebox[2]{#2}%
  \@ifundefined{ifGPcolor}{%
    \newif\ifGPcolor
    \GPcolorfalse
  }{}%
  \@ifundefined{ifGPblacktext}{%
    \newif\ifGPblacktext
    \GPblacktexttrue
  }{}%
  \let\gplgaddtomacro\g@addto@macro
  \gdef\gplbacktext{}%
  \gdef\gplfronttext{}%
  \makeatother
  \ifGPblacktext
    \def\colorrgb#1{}%
    \def\colorgray#1{}%
  \else
    \ifGPcolor
      \def\colorrgb#1{\color[rgb]{#1}}%
      \def\colorgray#1{\color[gray]{#1}}%
      \expandafter\def\csname LTw\endcsname{\color{white}}%
      \expandafter\def\csname LTb\endcsname{\color{black}}%
      \expandafter\def\csname LTa\endcsname{\color{black}}%
      \expandafter\def\csname LT0\endcsname{\color[rgb]{1,0,0}}%
      \expandafter\def\csname LT1\endcsname{\color[rgb]{0,1,0}}%
      \expandafter\def\csname LT2\endcsname{\color[rgb]{0,0,1}}%
      \expandafter\def\csname LT3\endcsname{\color[rgb]{1,0,1}}%
      \expandafter\def\csname LT4\endcsname{\color[rgb]{0,1,1}}%
      \expandafter\def\csname LT5\endcsname{\color[rgb]{1,1,0}}%
      \expandafter\def\csname LT6\endcsname{\color[rgb]{0,0,0}}%
      \expandafter\def\csname LT7\endcsname{\color[rgb]{1,0.3,0}}%
      \expandafter\def\csname LT8\endcsname{\color[rgb]{0.5,0.5,0.5}}%
    \else
      \def\colorrgb#1{\color{black}}%
      \def\colorgray#1{\color[gray]{#1}}%
      \expandafter\def\csname LTw\endcsname{\color{white}}%
      \expandafter\def\csname LTb\endcsname{\color{black}}%
      \expandafter\def\csname LTa\endcsname{\color{black}}%
      \expandafter\def\csname LT0\endcsname{\color{black}}%
      \expandafter\def\csname LT1\endcsname{\color{black}}%
      \expandafter\def\csname LT2\endcsname{\color{black}}%
      \expandafter\def\csname LT3\endcsname{\color{black}}%
      \expandafter\def\csname LT4\endcsname{\color{black}}%
      \expandafter\def\csname LT5\endcsname{\color{black}}%
      \expandafter\def\csname LT6\endcsname{\color{black}}%
      \expandafter\def\csname LT7\endcsname{\color{black}}%
      \expandafter\def\csname LT8\endcsname{\color{black}}%
    \fi
  \fi
    \setlength{\unitlength}{0.0500bp}%
    \ifx\gptboxheight\undefined%
      \newlength{\gptboxheight}%
      \newlength{\gptboxwidth}%
      \newsavebox{\gptboxtext}%
    \fi%
    \setlength{\fboxrule}{0.5pt}%
    \setlength{\fboxsep}{1pt}%
\begin{picture}(4320.00,3024.00)%
    \gplgaddtomacro\gplbacktext{%
      \csname LTb\endcsname
      \put(682,1104){\makebox(0,0)[r]{\strut{}$-2$}}%
      \put(682,1783){\makebox(0,0)[r]{\strut{}$0$}}%
      \put(682,2463){\makebox(0,0)[r]{\strut{}$2$}}%
      \put(814,374){\makebox(0,0){\strut{}$-2$}}%
      \put(1591,374){\makebox(0,0){\strut{}$-1$}}%
      \put(2369,374){\makebox(0,0){\strut{}$0$}}%
      \put(3146,374){\makebox(0,0){\strut{}$1$}}%
      \put(3923,374){\makebox(0,0){\strut{}$2$}}%
      \csname LTb\endcsname
      \put(2563,900){\makebox(0,0)[l]{\strut{}IR}}%
    }%
    \gplgaddtomacro\gplfronttext{%
      \csname LTb\endcsname
      \put(237,1698){\rotatebox{-270}{\makebox(0,0){\strut{}$\mathcal{V}_{\rm NESS}$}}}%
      \put(2368,154){\makebox(0,0){\strut{}$\varphi$}}%
      \put(2368,2693){\makebox(0,0){\strut{}}}%
      \csname LTb\endcsname
      \put(3068,2591){\makebox(0,0)[r]{\strut{}(b)   $\hspace{3ex} T < T_{\rm c}\hspace{1em} $  NESS}}%
      \csname LTb\endcsname
      \put(3068,2371){\makebox(0,0)[r]{\strut{}EQ}}%
    }%
    \gplbacktext
    \put(0,0){\includegraphics{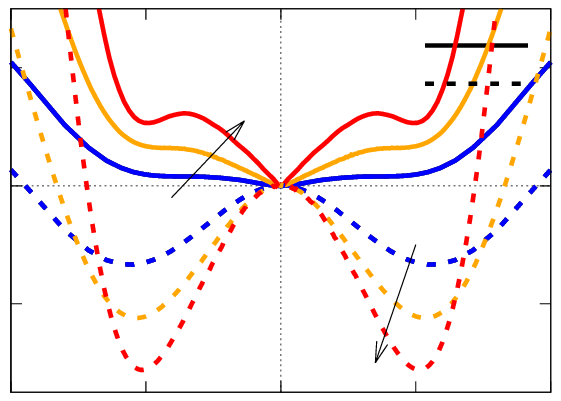}}%
    \gplfronttext
  \end{picture}%
\endgroup

%% file: Grad_LT_d4.tex
\begingroup
  \makeatletter
  \providecommand\color[2][]{%
    \GenericError{(gnuplot) \space\space\space\@spaces}{%
      Package color not loaded in conjunction with
      terminal option `colourtext'%
    }{See the gnuplot documentation for explanation.%
    }{Either use 'blacktext' in gnuplot or load the package
      color.sty in LaTeX.}%
    \renewcommand\color[2][]{}%
  }%
  \providecommand\includegraphics[2][]{%
    \GenericError{(gnuplot) \space\space\space\@spaces}{%
      Package graphicx or graphics not loaded%
    }{See the gnuplot documentation for explanation.%
    }{The gnuplot epslatex terminal needs graphicx.sty or graphics.sty.}%
    \renewcommand\includegraphics[2][]{}%
  }%
  \providecommand\rotatebox[2]{#2}%
  \@ifundefined{ifGPcolor}{%
    \newif\ifGPcolor
    \GPcolorfalse
  }{}%
  \@ifundefined{ifGPblacktext}{%
    \newif\ifGPblacktext
    \GPblacktexttrue
  }{}%
  \let\gplgaddtomacro\g@addto@macro
  \gdef\gplbacktext{}%
  \gdef\gplfronttext{}%
  \makeatother
  \ifGPblacktext
    \def\colorrgb#1{}%
    \def\colorgray#1{}%
  \else
    \ifGPcolor
      \def\colorrgb#1{\color[rgb]{#1}}%
      \def\colorgray#1{\color[gray]{#1}}%
      \expandafter\def\csname LTw\endcsname{\color{white}}%
      \expandafter\def\csname LTb\endcsname{\color{black}}%
      \expandafter\def\csname LTa\endcsname{\color{black}}%
      \expandafter\def\csname LT0\endcsname{\color[rgb]{1,0,0}}%
      \expandafter\def\csname LT1\endcsname{\color[rgb]{0,1,0}}%
      \expandafter\def\csname LT2\endcsname{\color[rgb]{0,0,1}}%
      \expandafter\def\csname LT3\endcsname{\color[rgb]{1,0,1}}%
      \expandafter\def\csname LT4\endcsname{\color[rgb]{0,1,1}}%
      \expandafter\def\csname LT5\endcsname{\color[rgb]{1,1,0}}%
      \expandafter\def\csname LT6\endcsname{\color[rgb]{0,0,0}}%
      \expandafter\def\csname LT7\endcsname{\color[rgb]{1,0.3,0}}%
      \expandafter\def\csname LT8\endcsname{\color[rgb]{0.5,0.5,0.5}}%
    \else
      \def\colorrgb#1{\color{black}}%
      \def\colorgray#1{\color[gray]{#1}}%
      \expandafter\def\csname LTw\endcsname{\color{white}}%
      \expandafter\def\csname LTb\endcsname{\color{black}}%
      \expandafter\def\csname LTa\endcsname{\color{black}}%
      \expandafter\def\csname LT0\endcsname{\color{black}}%
      \expandafter\def\csname LT1\endcsname{\color{black}}%
      \expandafter\def\csname LT2\endcsname{\color{black}}%
      \expandafter\def\csname LT3\endcsname{\color{black}}%
      \expandafter\def\csname LT4\endcsname{\color{black}}%
      \expandafter\def\csname LT5\endcsname{\color{black}}%
      \expandafter\def\csname LT6\endcsname{\color{black}}%
      \expandafter\def\csname LT7\endcsname{\color{black}}%
      \expandafter\def\csname LT8\endcsname{\color{black}}%
    \fi
  \fi
    \setlength{\unitlength}{0.0500bp}%
    \ifx\gptboxheight\undefined%
      \newlength{\gptboxheight}%
      \newlength{\gptboxwidth}%
      \newsavebox{\gptboxtext}%
    \fi%
    \setlength{\fboxrule}{0.5pt}%
    \setlength{\fboxsep}{1pt}%
\begin{picture}(4320.00,3024.00)%
    \gplgaddtomacro\gplbacktext{%
      \csname LTb\endcsname
      \put(550,717){\makebox(0,0)[r]{\strut{}$0$}}%
      \put(550,1699){\makebox(0,0)[r]{\strut{}$4$}}%
      \put(550,2680){\makebox(0,0)[r]{\strut{}$8$}}%
      \put(1222,374){\makebox(0,0){\strut{}$-1$}}%
      \put(2303,374){\makebox(0,0){\strut{}$0$}}%
      \put(3383,374){\makebox(0,0){\strut{}$1$}}%
      \csname LTb\endcsname
      \put(898,962){\makebox(0,0)[l]{\strut{}$T < T_{\rm c}$}}%
    }%
    \gplgaddtomacro\gplfronttext{%
      \csname LTb\endcsname
      \put(237,1698){\rotatebox{-270}{\makebox(0,0){\strut{}$\mathcal{K}_{\rm NESS}$}}}%
      \put(2302,154){\makebox(0,0){\strut{}$\nabla \varphi$}}%
      \put(2302,2693){\makebox(0,0){\strut{}}}%
      \csname LTb\endcsname
      \put(2936,2630){\makebox(0,0)[r]{\strut{}(a)   $\hspace{3ex} d = 4\hspace{6ex} r=0$}}%
      \csname LTb\endcsname
      \put(2936,2410){\makebox(0,0)[r]{\strut{}$r=0.005$}}%
      \csname LTb\endcsname
      \put(2936,2190){\makebox(0,0)[r]{\strut{}$r=0.05$}}%
    }%
    \gplbacktext
    \put(0,0){\includegraphics{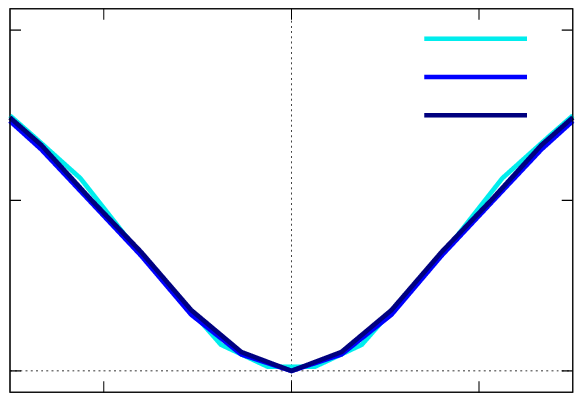}}%
    \gplfronttext
  \end{picture}%
\endgroup

%% file: Grad_LT_d2.tex
\begingroup
  \makeatletter
  \providecommand\color[2][]{%
    \GenericError{(gnuplot) \space\space\space\@spaces}{%
      Package color not loaded in conjunction with
      terminal option `colourtext'%
    }{See the gnuplot documentation for explanation.%
    }{Either use 'blacktext' in gnuplot or load the package
      color.sty in LaTeX.}%
    \renewcommand\color[2][]{}%
  }%
  \providecommand\includegraphics[2][]{%
    \GenericError{(gnuplot) \space\space\space\@spaces}{%
      Package graphicx or graphics not loaded%
    }{See the gnuplot documentation for explanation.%
    }{The gnuplot epslatex terminal needs graphicx.sty or graphics.sty.}%
    \renewcommand\includegraphics[2][]{}%
  }%
  \providecommand\rotatebox[2]{#2}%
  \@ifundefined{ifGPcolor}{%
    \newif\ifGPcolor
    \GPcolorfalse
  }{}%
  \@ifundefined{ifGPblacktext}{%
    \newif\ifGPblacktext
    \GPblacktexttrue
  }{}%
  \let\gplgaddtomacro\g@addto@macro
  \gdef\gplbacktext{}%
  \gdef\gplfronttext{}%
  \makeatother
  \ifGPblacktext
    \def\colorrgb#1{}%
    \def\colorgray#1{}%
  \else
    \ifGPcolor
      \def\colorrgb#1{\color[rgb]{#1}}%
      \def\colorgray#1{\color[gray]{#1}}%
      \expandafter\def\csname LTw\endcsname{\color{white}}%
      \expandafter\def\csname LTb\endcsname{\color{black}}%
      \expandafter\def\csname LTa\endcsname{\color{black}}%
      \expandafter\def\csname LT0\endcsname{\color[rgb]{1,0,0}}%
      \expandafter\def\csname LT1\endcsname{\color[rgb]{0,1,0}}%
      \expandafter\def\csname LT2\endcsname{\color[rgb]{0,0,1}}%
      \expandafter\def\csname LT3\endcsname{\color[rgb]{1,0,1}}%
      \expandafter\def\csname LT4\endcsname{\color[rgb]{0,1,1}}%
      \expandafter\def\csname LT5\endcsname{\color[rgb]{1,1,0}}%
      \expandafter\def\csname LT6\endcsname{\color[rgb]{0,0,0}}%
      \expandafter\def\csname LT7\endcsname{\color[rgb]{1,0.3,0}}%
      \expandafter\def\csname LT8\endcsname{\color[rgb]{0.5,0.5,0.5}}%
    \else
      \def\colorrgb#1{\color{black}}%
      \def\colorgray#1{\color[gray]{#1}}%
      \expandafter\def\csname LTw\endcsname{\color{white}}%
      \expandafter\def\csname LTb\endcsname{\color{black}}%
      \expandafter\def\csname LTa\endcsname{\color{black}}%
      \expandafter\def\csname LT0\endcsname{\color{black}}%
      \expandafter\def\csname LT1\endcsname{\color{black}}%
      \expandafter\def\csname LT2\endcsname{\color{black}}%
      \expandafter\def\csname LT3\endcsname{\color{black}}%
      \expandafter\def\csname LT4\endcsname{\color{black}}%
      \expandafter\def\csname LT5\endcsname{\color{black}}%
      \expandafter\def\csname LT6\endcsname{\color{black}}%
      \expandafter\def\csname LT7\endcsname{\color{black}}%
      \expandafter\def\csname LT8\endcsname{\color{black}}%
    \fi
  \fi
    \setlength{\unitlength}{0.0500bp}%
    \ifx\gptboxheight\undefined%
      \newlength{\gptboxheight}%
      \newlength{\gptboxwidth}%
      \newsavebox{\gptboxtext}%
    \fi%
    \setlength{\fboxrule}{0.5pt}%
    \setlength{\fboxsep}{1pt}%
\begin{picture}(4320.00,3024.00)%
    \gplgaddtomacro\gplbacktext{%
      \csname LTb\endcsname
      \put(550,717){\makebox(0,0)[r]{\strut{}$0$}}%
      \put(550,1699){\makebox(0,0)[r]{\strut{}$4$}}%
      \put(550,2680){\makebox(0,0)[r]{\strut{}$8$}}%
      \put(1222,374){\makebox(0,0){\strut{}$-1$}}%
      \put(2303,374){\makebox(0,0){\strut{}$0$}}%
      \put(3383,374){\makebox(0,0){\strut{}$1$}}%
      \csname LTb\endcsname
      \put(898,962){\makebox(0,0)[l]{\strut{}$T < T_{\rm c}$}}%
    }%
    \gplgaddtomacro\gplfronttext{%
      \csname LTb\endcsname
      \put(237,1698){\rotatebox{-270}{\makebox(0,0){\strut{}$\mathcal{K}_{\rm NESS}$}}}%
      \put(2302,154){\makebox(0,0){\strut{}$\nabla \varphi$}}%
      \put(2302,2693){\makebox(0,0){\strut{}}}%
      \csname LTb\endcsname
      \put(2936,2630){\makebox(0,0)[r]{\strut{}(b)   $\hspace{2ex}d = 2 \hspace{3ex} r=0.005$}}%
      \csname LTb\endcsname
      \put(2936,2410){\makebox(0,0)[r]{\strut{}$r=0.05$}}%
      \csname LTb\endcsname
      \put(2936,2190){\makebox(0,0)[r]{\strut{}$r=0.1$}}%
    }%
    \gplbacktext
    \put(0,0){\includegraphics{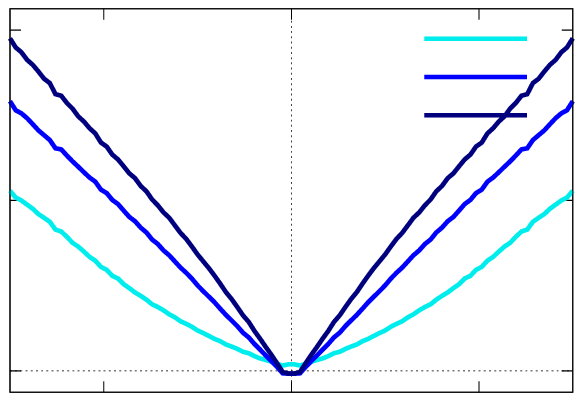}}%
    \gplfronttext
  \end{picture}%
\endgroup

%% file: fits.tex
\begingroup
  \makeatletter
  \providecommand\color[2][]{%
    \GenericError{(gnuplot) \space\space\space\@spaces}{%
      Package color not loaded in conjunction with
      terminal option `colourtext'%
    }{See the gnuplot documentation for explanation.%
    }{Either use 'blacktext' in gnuplot or load the package
      color.sty in LaTeX.}%
    \renewcommand\color[2][]{}%
  }%
  \providecommand\includegraphics[2][]{%
    \GenericError{(gnuplot) \space\space\space\@spaces}{%
      Package graphicx or graphics not loaded%
    }{See the gnuplot documentation for explanation.%
    }{The gnuplot epslatex terminal needs graphicx.sty or graphics.sty.}%
    \renewcommand\includegraphics[2][]{}%
  }%
  \providecommand\rotatebox[2]{#2}%
  \@ifundefined{ifGPcolor}{%
    \newif\ifGPcolor
    \GPcolorfalse
  }{}%
  \@ifundefined{ifGPblacktext}{%
    \newif\ifGPblacktext
    \GPblacktexttrue
  }{}%
  \let\gplgaddtomacro\g@addto@macro
  \gdef\gplbacktext{}%
  \gdef\gplfronttext{}%
  \makeatother
  \ifGPblacktext
    \def\colorrgb#1{}%
    \def\colorgray#1{}%
  \else
    \ifGPcolor
      \def\colorrgb#1{\color[rgb]{#1}}%
      \def\colorgray#1{\color[gray]{#1}}%
      \expandafter\def\csname LTw\endcsname{\color{white}}%
      \expandafter\def\csname LTb\endcsname{\color{black}}%
      \expandafter\def\csname LTa\endcsname{\color{black}}%
      \expandafter\def\csname LT0\endcsname{\color[rgb]{1,0,0}}%
      \expandafter\def\csname LT1\endcsname{\color[rgb]{0,1,0}}%
      \expandafter\def\csname LT2\endcsname{\color[rgb]{0,0,1}}%
      \expandafter\def\csname LT3\endcsname{\color[rgb]{1,0,1}}%
      \expandafter\def\csname LT4\endcsname{\color[rgb]{0,1,1}}%
      \expandafter\def\csname LT5\endcsname{\color[rgb]{1,1,0}}%
      \expandafter\def\csname LT6\endcsname{\color[rgb]{0,0,0}}%
      \expandafter\def\csname LT7\endcsname{\color[rgb]{1,0.3,0}}%
      \expandafter\def\csname LT8\endcsname{\color[rgb]{0.5,0.5,0.5}}%
    \else
      \def\colorrgb#1{\color{black}}%
      \def\colorgray#1{\color[gray]{#1}}%
      \expandafter\def\csname LTw\endcsname{\color{white}}%
      \expandafter\def\csname LTb\endcsname{\color{black}}%
      \expandafter\def\csname LTa\endcsname{\color{black}}%
      \expandafter\def\csname LT0\endcsname{\color{black}}%
      \expandafter\def\csname LT1\endcsname{\color{black}}%
      \expandafter\def\csname LT2\endcsname{\color{black}}%
      \expandafter\def\csname LT3\endcsname{\color{black}}%
      \expandafter\def\csname LT4\endcsname{\color{black}}%
      \expandafter\def\csname LT5\endcsname{\color{black}}%
      \expandafter\def\csname LT6\endcsname{\color{black}}%
      \expandafter\def\csname LT7\endcsname{\color{black}}%
      \expandafter\def\csname LT8\endcsname{\color{black}}%
    \fi
  \fi
    \setlength{\unitlength}{0.0500bp}%
    \ifx\gptboxheight\undefined%
      \newlength{\gptboxheight}%
      \newlength{\gptboxwidth}%
      \newsavebox{\gptboxtext}%
    \fi%
    \setlength{\fboxrule}{0.5pt}%
    \setlength{\fboxsep}{1pt}%
\begin{picture}(7200.00,3528.00)%
    \gplgaddtomacro\gplbacktext{%
      \csname LTb\endcsname
      \put(550,682){\makebox(0,0)[r]{\strut{}$0$}}%
      \put(550,2432){\makebox(0,0)[r]{\strut{}$1$}}%
      \put(960,374){\makebox(0,0){\strut{}$-0.5$}}%
      \put(3743,374){\makebox(0,0){\strut{}$0$}}%
      \put(6525,374){\makebox(0,0){\strut{}$0.5$}}%
      \csname LTb\endcsname
      \put(4744,-192){\makebox(0,0)[l]{\strut{}IR}}%
    }%
    \gplgaddtomacro\gplfronttext{%
      \csname LTb\endcsname
      \put(237,1950){\rotatebox{-270}{\makebox(0,0){\strut{}$\mathcal{V}_{\rm NESS}$}}}%
      \put(3742,154){\makebox(0,0){\strut{}$\varphi$}}%
      \put(3742,3197){\makebox(0,0){\strut{}}}%
      \csname LTb\endcsname
      \put(5058,3109){\makebox(0,0)[r]{\strut{}NESS Monte Carlo}}%
      \csname LTb\endcsname
      \put(5058,2889){\makebox(0,0)[r]{\strut{}fit with $c_1 |\varphi| + a_2 \varphi^2 + c_3 |\varphi|^3 + a_4 \varphi^4$}}%
      \csname LTb\endcsname
      \put(5058,2669){\makebox(0,0)[r]{\strut{}fit with $ a_2 \varphi^2 + a_4 \varphi^4 +  a_6 \varphi^6 + a_8 \varphi^8$}}%
    }%
    \gplbacktext
    \put(0,0){\includegraphics{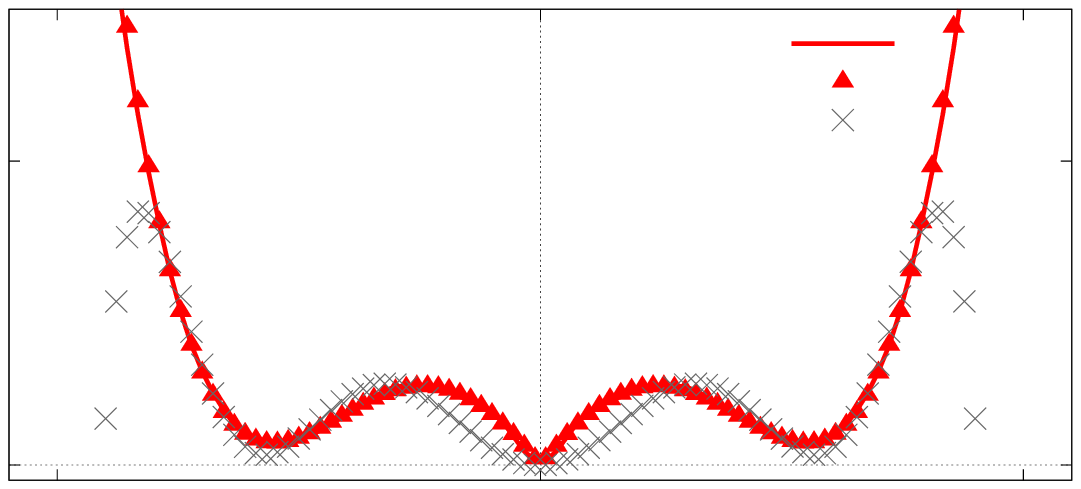}}%
    \gplfronttext
  \end{picture}%
\endgroup

%% file: Mag_t.tex
\begingroup
  \makeatletter
  \providecommand\color[2][]{%
    \GenericError{(gnuplot) \space\space\space\@spaces}{%
      Package color not loaded in conjunction with
      terminal option `colourtext'%
    }{See the gnuplot documentation for explanation.%
    }{Either use 'blacktext' in gnuplot or load the package
      color.sty in LaTeX.}%
    \renewcommand\color[2][]{}%
  }%
  \providecommand\includegraphics[2][]{%
    \GenericError{(gnuplot) \space\space\space\@spaces}{%
      Package graphicx or graphics not loaded%
    }{See the gnuplot documentation for explanation.%
    }{The gnuplot epslatex terminal needs graphicx.sty or graphics.sty.}%
    \renewcommand\includegraphics[2][]{}%
  }%
  \providecommand\rotatebox[2]{#2}%
  \@ifundefined{ifGPcolor}{%
    \newif\ifGPcolor
    \GPcolorfalse
  }{}%
  \@ifundefined{ifGPblacktext}{%
    \newif\ifGPblacktext
    \GPblacktexttrue
  }{}%
  \let\gplgaddtomacro\g@addto@macro
  \gdef\gplbacktext{}%
  \gdef\gplfronttext{}%
  \makeatother
  \ifGPblacktext
    \def\colorrgb#1{}%
    \def\colorgray#1{}%
  \else
    \ifGPcolor
      \def\colorrgb#1{\color[rgb]{#1}}%
      \def\colorgray#1{\color[gray]{#1}}%
      \expandafter\def\csname LTw\endcsname{\color{white}}%
      \expandafter\def\csname LTb\endcsname{\color{black}}%
      \expandafter\def\csname LTa\endcsname{\color{black}}%
      \expandafter\def\csname LT0\endcsname{\color[rgb]{1,0,0}}%
      \expandafter\def\csname LT1\endcsname{\color[rgb]{0,1,0}}%
      \expandafter\def\csname LT2\endcsname{\color[rgb]{0,0,1}}%
      \expandafter\def\csname LT3\endcsname{\color[rgb]{1,0,1}}%
      \expandafter\def\csname LT4\endcsname{\color[rgb]{0,1,1}}%
      \expandafter\def\csname LT5\endcsname{\color[rgb]{1,1,0}}%
      \expandafter\def\csname LT6\endcsname{\color[rgb]{0,0,0}}%
      \expandafter\def\csname LT7\endcsname{\color[rgb]{1,0.3,0}}%
      \expandafter\def\csname LT8\endcsname{\color[rgb]{0.5,0.5,0.5}}%
    \else
      \def\colorrgb#1{\color{black}}%
      \def\colorgray#1{\color[gray]{#1}}%
      \expandafter\def\csname LTw\endcsname{\color{white}}%
      \expandafter\def\csname LTb\endcsname{\color{black}}%
      \expandafter\def\csname LTa\endcsname{\color{black}}%
      \expandafter\def\csname LT0\endcsname{\color{black}}%
      \expandafter\def\csname LT1\endcsname{\color{black}}%
      \expandafter\def\csname LT2\endcsname{\color{black}}%
      \expandafter\def\csname LT3\endcsname{\color{black}}%
      \expandafter\def\csname LT4\endcsname{\color{black}}%
      \expandafter\def\csname LT5\endcsname{\color{black}}%
      \expandafter\def\csname LT6\endcsname{\color{black}}%
      \expandafter\def\csname LT7\endcsname{\color{black}}%
      \expandafter\def\csname LT8\endcsname{\color{black}}%
    \fi
  \fi
    \setlength{\unitlength}{0.0500bp}%
    \ifx\gptboxheight\undefined%
      \newlength{\gptboxheight}%
      \newlength{\gptboxwidth}%
      \newsavebox{\gptboxtext}%
    \fi%
    \setlength{\fboxrule}{0.5pt}%
    \setlength{\fboxsep}{1pt}%
\begin{picture}(7200.00,3528.00)%
    \gplgaddtomacro\gplbacktext{%
      \csname LTb\endcsname
      \put(814,594){\makebox(0,0)[r]{\strut{}$0$}}%
      \put(814,1498){\makebox(0,0)[r]{\strut{}$0.1$}}%
      \put(814,2403){\makebox(0,0)[r]{\strut{}$0.2$}}%
      \put(814,3307){\makebox(0,0)[r]{\strut{}$0.3$}}%
      \put(946,374){\makebox(0,0){\strut{}$1$}}%
      \put(2099,374){\makebox(0,0){\strut{}$50$}}%
      \put(3275,374){\makebox(0,0){\strut{}$100$}}%
      \put(4451,374){\makebox(0,0){\strut{}$150$}}%
      \put(5627,374){\makebox(0,0){\strut{}$200$}}%
      \put(6803,374){\makebox(0,0){\strut{}$250$}}%
    }%
    \gplgaddtomacro\gplfronttext{%
      \csname LTb\endcsname
      \put(185,1950){\rotatebox{-270}{\makebox(0,0){\strut{}$\langle |\varphi| \rangle$}}}%
      \put(3874,154){\makebox(0,0){\strut{}$t$}}%
      \put(3874,3197){\makebox(0,0){\strut{}}}%
      \csname LTb\endcsname
      \put(5948,2564){\makebox(0,0)[r]{\strut{}Monte Carlo}}%
      \csname LTb\endcsname
      \put(5948,2344){\makebox(0,0)[r]{\strut{}Mean Field}}%
    }%
    \gplbacktext
    \put(0,0){\includegraphics{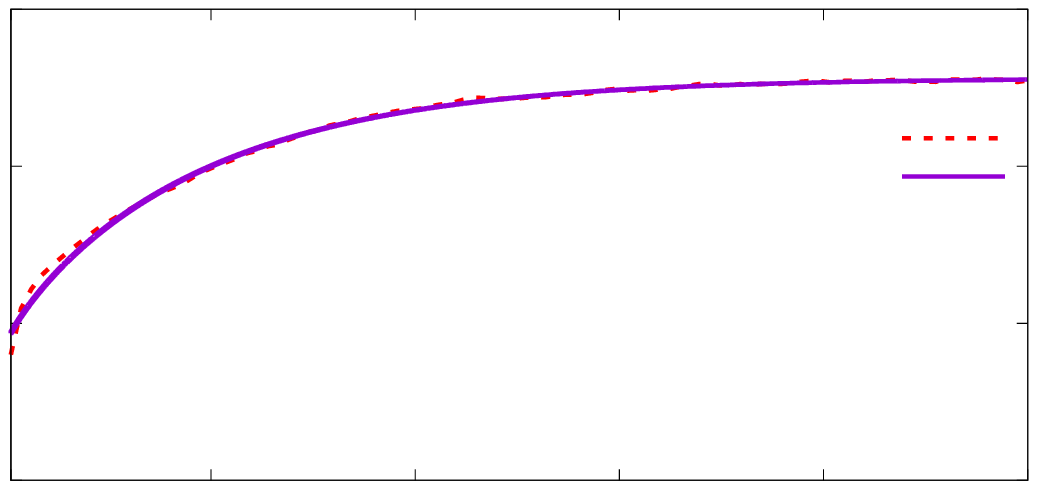}}%
    \gplfronttext
  \end{picture}%
\endgroup